\documentclass[preprint,aps,superscriptaddress,showkeys,nofootinbib,11pt]{revtex4}
\pdfoutput=1
\usepackage{graphicx}
\usepackage{amsmath}
\usepackage{amsfonts}
\usepackage{amssymb}
\usepackage{xcolor,soul}
\usepackage{subfigure}
\usepackage{pdfpages}
\usepackage{longtable}
\usepackage{array,booktabs,enumitem}
\newcolumntype{P}[1]{>{\endgraf\vspace*{-\baselineskip}}p{#1}}

\definecolor{LightGray}{gray}{0.85}
\usepackage{hyperref}

\newcommand{\hyphenB}[1]{{\operatorname{\mathit{#1}}}}
\def\beq{\begin{equation}}
\def\eeq{\end{equation}}
\def\bea{\begin{eqnarray}}
\def\eea{\end{eqnarray}}
\def\be{\begin{equation}}
\def\ee{\end{equation}}

\def\nno{\nonumber}
\def\bse{\begin{subequations}}
	\def\ese{\end{subequations}}

\definecolor{LightGray}{gray}{0.85}

\graphicspath{{./figs/}}
\begin{document}

\title{Connecting CMB anisotropy and cold dark matter phenomenology via reheating}

\author{Debaprasad Maity}
 \email{debu@iitg.ac.in}
\author{Pankaj Saha}%
 \email{pankaj.saha@iitg.ac.in}
\affiliation{%
Department of Physics, Indian Institute of Technology Guwahati.\\
 Guwahati, Assam, India 
}%

\begin{abstract}
Understanding the properties of dark matter has proved to be one of the most challenging problems of particle phenomenology. In this paper, we have tried to understand the phenomenology of dark matter in light of very well understood properties of cosmic microwave background (CMB) anisotropy. To connect these two, inflation and its subsequent evolution known as the reheating phase play the important role. Following the previous analysis, we first established one-to-one correspondence between the CMB power spectrum and the reheating temperature assuming the perturbative reheating scenario. Further by incorporating a possible dark matter candidate through the radiation annihilation process during reheating and the current value of dark matter abundance, we constrain the dark matter parameter space through the inflationary power spectrum for different inflationary models.   

\end{abstract}
\maketitle

\tableofcontents

\section{Introduction}
The inflationary paradigm\cite{guth, linde1, steinhardt} was introduced initially to solve the initial condition problem of the hot big bang model of standard cosmology. Associated with the inflation, the generic inflation energy scale is assumed to be $\geq 10^{10}{\rm GeV}$. On the other hand, successful big bang nucleosynthesis(BBN) predicting the current light elements abundance requires our universe to be radiation dominated with the minimum temperature to be $T_{\rm BBN}\sim 1$ MeV\cite{Kawasaki:1999na, Kawasaki:2000en, Steigman:2007xt, Fields:2014uja}. 
Therefore, the evolution of our universe from the inflation to BBN needs highly nontrivial dynamics which not only produces all the matter particles we see today but also connects these widely separated energy scales through the complex non-linear process and thermalization. 
 
The signature of the inflationary evolution can be extracted from the cosmic microwave background (CMB) measurements\cite{encyclopedia, PLANCK}. On the other hand, the BBN is very successful in explaining the light element abundance in the present universe. However, until now the period between the aforementioned two cosmological eras is poorly understood. One of the reasons is our observational limitations to directly probe this phase, and most importantly the dynamics during this phase are expected to be highly nonlinear in nature as just noted above. 
This phase which has been dubbed as the reheating era \cite{Albrecht:1982mp,Dolgov:1982th,Abbott:1982hn,Traschen:1990sw} is, in general, parametrized by reheating temperature $T_{\rm re}$ which is defined at the instant when the inflaton decay rate becomes equal to the expansion rate of the universe during reheating.
After the reheating period is over, the reheating temperature can be directly connected with the current CMB temperature through background expansion. Therefore, it is possible to constrain the inflationary models through the subsequent reheating phase and CMB anisotropy\cite{Dai:2014jja,Creminelli:2014fca,Martin:2014nya}. This idea of the reheating constraint on inflation dynamics has recently been studied extensively for various inflationary models\cite{Cook:2015vqa,Ellis:2015pla,Ueno:2016dim,Eshaghi:2016kne,DiMarco:2017zek,Bhattacharya:2017ysa,Drewes:2017fmn}.

One of the important assumptions of the aforementioned reheating constraint analysis is that during the reheating phase, the inflaton decays only into the radiation component. Therefore, it has an inherent limitation to extend the analysis beyond radiation. In this paper, our main goal is to extend and generalize the existing analysis of reheating constraints considering the effect of dark matter production during the reheating phase. 
In the current epoch, apart from the cosmological constant, dark matter and CMB are the two main components of our universe. From the observational point of view CMB is the most powerful probe to understand the evolution of the universe.
Through CMB, we not only understand the background expansion of our universe but also understand various physical processes acting during the formation of a large-scale structure we see today. 
Dark matter is believed to play one of the important roles in the aforementioned processes of structure formation.
However, because of very weak interaction with the visible matter field, dark matter is very difficult to detect. From the background evolution, we only know our universe to be $23\%$ dark matter dominant out of the total energy budget of the universe. 
This fact motivates us to understand the following question: {\it does the CMB have any role to play in understanding the dark matter phenomenology?}

To answer this question, we think it is the reheating phase that has the potential to shed some light on the possible connection between the CMB and the current dark matter abundance. With this in mind and following our previous work \cite{Maity:2017thw}, we assume decaying dynamics of the inflaton to be perturbative during reheating and dark matter is produced through annihilation of the radiation component. Inflation decaying into various fields and their observable effects has already been extensively studied before \cite{Chung:1998rq,Giudice:2000ex,Allahverdi:2002nb,Allahverdi:2002pu,Pallis:2004yy,Kane:2015qea, Maity:2016uyn}. However, as already emphasized, our main goal is to connect the dark matter phenomenology and CMB anisotropy via inflation and reheating. Therefore our analysis will be an important generalization of the previous work \cite{Dai:2014jja}. 

Since inflaton is decaying through a perturbative process, the assumption of a complete conversion of inflation into radiation at the instant of reheating will not hold which has been extensively considered before. This assumption is applicable if the reheating is instantaneous. But in general, this is not the case. Therefore, we will see that there will be a significant correction in the reheating temperature as only a fraction of total inflaton energy is converted into radiation at the time \cite{Giudice:2000ex}  when $\Gamma_{\phi}=H$. For simplicity, we will assume that the dark matter is produced only through an annihilation channel from the radiation component. 
We believe our study can also help us gain more insight into the production mechanism of dark matter intimately tied with the inflationary and reheating dynamics. We leave explicit model construction for our future studies.

To this end let us point out an important observation we made through our analysis. The production of dark matter particle in an expanding universe such as ours generally can be of two types. Depending upon the initial energy density and the
 rate of background expansion, if the annihilation cross section to dark matter is large, the produced particle will reach thermal equilibrium before freeze-out to current abundance \cite{Griest:1990kh,Edsjo:1997bg,DAgnolo:2015ujb,Chang:2009yt,Fan:2010gt,Fitzpatrick:2012ix,Cirelli:2008pk,ArkaniHamed:2008qn,Pospelov:2008jd,Fox:2008kb}, which is the well-known ``freeze-out" mechanism. On the other hand, if the annihilation cross section is small enough, the comoving dark matter particle density becomes constant much before it can reach thermal equilibrium with the background radiation. This production mechanism is known as the ``freeze-in'' mechanism.  
 In the particle physics context, the existing model of this type is known as feebly interacting dark matter \cite{Hall:2009bx,Tenkanen:2016twd,Heikinheimo:2016hid,Bernal:2017kxu}. Interestingly, if we consider the reheating process to be perturbative, our analysis shows that for dark matter mass much larger than the reheating temperature, the current dark matter abundance can be produced only via the freeze-in mechanism. The reason is the unique boundary conditions set by the inflation. However, for dark matter mass smaller than the reheating temperature, both mechanisms will work. For the present purpose, we have explicitly considered the freeze-in mechanism. A detailed analysis of different mechanisms will be studied elsewhere. 

 The remainder of this work is organized as follows. In the first two sections, we essentially review the well-known results to set the stage for our current analysis. In Sec. 2, we will discuss the inflationary observables and its connection with CMB. In Sec. 3, we describe the set of Boltzmann equations that describes the dynamics of the reheating phase. As has been mentioned in the Introduction, we will calculate the reheating temperature and corresponding e-folding number considering the explicit decay of inflaton. For this, we solve the system of Boltzmann equations numerically and identify the individual components during reheating with their current abundance. With this identification, we are able to shed light on the dark matter through CMB anisotropy. We study different inflationary models and their constraints on the dark matter phenomenology. Finally, we conclude in Sec. 5.
 
\section{Inflationary observables connection with CMB}
One of the important observables in CMB is the correlation of temperature fluctuations which is directly related to the inflationary observable known as scalar spectral index $n_s$. The equations governing the dynamics of the aforementioned scalar field called inflaton $\phi$ with a potential $V(\phi)$ is 
\begin{eqnarray}
\ddot{\phi} + (3H + \Gamma_{\phi})\dot{\phi} +  V'(\phi)=0,\\
H^2 =\left(\frac{\dot a}{a}\right)^2= \frac{1}{3M_p^2}\rho_t ,
\label{EOM}
\end{eqnarray}
where, we consider the following Friedmann-Roberson-Walker(FRW) spacetime background $ds^2 = -dt^2 + a(t)^2 (dx^2 + dy^2 +dz^2)$.
$H$ is the Hubble expansion rate and $M_p(=1/\sqrt{8\pi G})$ is the effective Planck mass. In this paper, we will discuss our results based on the canonical scalar field models. A More general model will be considered elsewhere. The decay term $\Gamma_{\phi}\dot{\phi}$ in the above equation is assumed to be negligible during inflation, however, it will become important during the reheating period. Therefore, during inflation total energy density of the universe will be dominated by the inflaton energy $\rho_t = \rho_{\phi}$. As is well known that almost homogeneous temperature $T_0 \simeq 2.7{\rm K}$ of the CMB can be shown to be intimately tied with the slow-roll nature of inflaton dynamics, and it is parametrized in terms of potential $V(\phi) $ as follows,
\begin{eqnarray}
 \epsilon = \frac{1}{2}M_p^2\left[\frac{V'(\phi)}{V(\phi)}\right]^2~~~\eta 
= M_p^2\left[\frac{V''(\phi)}{V(\phi)}\right] .
\end{eqnarray}
Once we define the background inflationary dynamics, the main  quantities of interest
are the amplitude of the inflaton fluctuation $A_s$, the spectral index, $n_s$, and the tensor-to-scalar
ratio, $r$, which in terms of the slow-roll parameters are, 
\begin{eqnarray}
n_s = 1 - 6\epsilon_k + 2\eta;~~~~~~r = 16\epsilon.
\end{eqnarray}
The CMB normalization and the temperature correlation are in one-to-one correspondence with $A_s$ and  $n_s$ respectively. Therefore, those observables are directly used to constrain the inflationary models. Tensor to scalar ratio $r$, which is related to the inflationary energy scale has its signature in the polarization $B$ mode of CMB, which has not yet been observed. All those quantities are defined for a particular cosmological scale $k$ which is the  pivot scale of CMB, $k/a_0 = 0.05 {\rm Mpc^{-1}}$. The latest bound on the scalar spectral index is \cite{PLANCK} given as $n_s =0.9659 \pm 0.0082$ for $\Lambda CDM+r$ model from Planck data alone or $n_s =0.9670 \pm 0.0074 $ from Planck and BK14 and BAO data. In our subsequent analysis, we will assign all the inflationary parameters at the aforementioned CMB scale at the time of its horizon crossing during inflation.  

Further, important inflationary quantities that will be considered are the Hubble parameter $H_k$ and e-folding number $N_k$ for a particular scale $k$ (CMB pivot scale) at its horizon crossing. Those quantities will be described in the appropriate places, but before that in the next section, we will review the Boltzmann equation for three different energy components namely inflaton, radiation, and dark matter.

\section{Dark Matter during reheating}
\subsection{Basic equations}
As has been emphasized in our previous discussions, the information of CMB has the potential to shed light on the dark matter sector through the reheating phase. Production of dark matter like particles considering different models and its phenomenology has  already been worked out in detail in the literature considering the decaying inflaton during reheating\cite{Chung:1998rq,Allahverdi:2002nb,Allahverdi:2002pu,Pallis:2004yy,Tenkanen:2016twd, Nurmi:2015ema,Bastero-Gil:2015lga, Heikinheimo:2016yds,Kainulainen:2016vzv,Visinelli:2017qga, Chen:2017kvz, Enqvist:2017kzh, DEramo:2017ecx}. Also, how a nonzero Higgs vacuum expectation value during inflation can impact on the standard reheating history of the universe has been discussed in\cite{Enqvist:2013kaa,Kusenko:2014lra,Freese:2017ace}. However the direct connection of the aforementioned analysis with the CMB has never been carefully looked into.
Therefore, combining the analysis mentioned in the previous section with the existing reheating analysis, in the subsequent sections, we will uncover a surprising connection between the CMB and dark matter phenomenology. Our study opens up a new avenue toward understanding the detail properties of the dark matter though CMB observations.
 
It is well known that after the end of inflation the universe becomes extremely homogeneous. Therefore, to set in the subsequent evolution, the inflaton field has to go through the reheating phase when it decays into other fields and radiation. Depending upon the coupling with the inflaton field, the reheating field can have either perturbative or nonperturbative production. For our current analysis, we will consider the purely the perturbative reheating process. 
Therefore, we essentially follow the existing analysis by considering the evolution of Boltzmann equations for three different energy components consisting of the inflation energy density $\rho_{\phi}$, the radiation energy density $\rho_{\phi}$ and the dark matter particle number density $n_X$\cite{Giudice:2000ex, book22}. 
\begin{eqnarray}
\frac{d \rho_\phi}{dt} &=& -3H(1+w_{\phi}) \rho_{\phi} - \Gamma_{\phi} (1+w_{\phi})\rho_\phi
\label{binp} \\
\frac{d \rho_R}{dt} &=& -4H \rho_R +\Gamma_\phi \rho_\phi +
\langle \sigma v \rangle 2\langle E_X \rangle 
\left[ n_X^2 - \left({n_{X,eq}}\right)^2\right]
\label{binr} \\
\frac{d n_X}{dt} &=& -3H n_X
-\langle \sigma v \rangle  \left[ n_X^2 -
\left({n_{X,eq}}\right)^2\right] ,
\label{Boltzmann1} \
\end{eqnarray}
and the background expansion is given by
\begin{equation} \label{Nreeq}
H^2=\frac{8\pi}{3M_{Pl}^2} (\rho_\phi +\rho_R +\rho_X) \ .
\end{equation}

where, $\left<E_X\right> = \rho_X/n_X  \simeq \sqrt{M_X^2 + (3T)^2}$ is the average energy density of a single component dark matter X particle and $n_X^{eq}$ is the equilibrium number density of the matter particle of mass $M_X$ at the equilibrium background temperature $T$. $\Gamma_{\phi}$ is the inflaton decay constant. As has been mentioned, the dark matter particles create and annihilate into radiation with a thermal-averaged cross section $\langle \sigma v\rangle$. $w_{\phi}$ is the average equation of state for an oscillating scalar field (inflaton)\cite{Mukhanov:2005sc}, 
\begin{equation}
 w_{\phi} = \frac{p_{\phi}}{\rho_{\phi}} \simeq \frac{\langle \phi V'(\phi) -2V(\phi)\rangle}{\langle \phi V'(\phi) + 2V(\phi)\rangle}
 \label{eos}
\end{equation}
For an inflaton potential $V(\phi) \propto \phi^n$, it is found to be $w_{\phi} = (n-2)/(n+2)$. At this point let us state an important difference of our work and that of \cite{Dai:2014jja,Drewes:2017fmn}. 
In those works, the equation of state parameters for the reheating period is expressed as that of an effective single fluid( comprising of inflaton and its decay products) equation of state. This is taken to be constant during the entire reheating period. In the present work, as we are explicitly solving the Boltzmann equations for different components of the universe during reheating, we need not consider the single field equation of the state parameter, but rather the quantity that is important here is the equation of state parameter for the homogeneous component of inflaton during oscillation. We will see that, for the models considered in the present work, the inflation equation of state is effectively given $w_{\phi}=0$. The general equation of state will have a considerable effect on the reheating state which we will consider in separate work. At this stage let us emphasize the fact that, nonperturbative decay could have a potential impact on our conclusion which we leave for our future studies.

Our goal of this paper is to look into a wide range of dark matter mass, $M_X$ which can be greater as well as less than the reheating temperature. We also assume the dark matter to follow the fermionic distribution having the internal degree of freedom $g$. Therefore, in thermal equilibrium the number density at temperature $T$ can be expressed as, 
\begin{equation}
n_{X,eq} = \frac{g}{2\pi^2} \int_{m_X}^{\infty} \frac{\sqrt{E^2 - M_X^2}}{e^{E/T} + 1} EdE \simeq
\frac{g T^3}{2\pi^2} \left(\frac{M_X}{T}\right)^2K_2\left(\frac{M_X}{T}\right) ,
\end{equation}
where, $K_2$ is the modified Bessel function of the second kind \cite{Giudice:2000ex}.

Now, in order to solve the equations numerically, it is convenient to work in terms of the following dimensionless quantities, 
\begin{eqnarray}
	\Phi \equiv \frac{\rho_{\phi}A^3}{m_{\phi}^4}, ~~~~~R \equiv \frac{\rho_R A^4}{m_{\phi}^4},~~~~X \equiv \frac{n_X A^3}{m_{\phi}^3} ,
\end{eqnarray}
The Boltzmann equations.\ref{Boltzmann1} in terms of these comoving dimensionless  variables are 
\begin{eqnarray}
	\frac{d\Phi}{dA} &=& -c_1 \frac{A^{1/2} }{\mathbb{H}}\Phi;\\
	\frac{dR}{dA} &=& c_1 \frac{A^{3/2}}{\mathbb{H}}\Phi + c_2 \frac{A^{-3/2} 2\left<E_X\right> \left<\sigma v\right> M_{pl}}{\mathbb{H}}\left(X^2 - X_{eq}^2\right);\\
	\frac{dX}{dA} &=& - c_2 \frac{A^{-5/2}  \left<\sigma v\right> m_{\phi}~M_{pl}}{\mathbb{H}}\left(X^2 - X_{eq}^2\right);
	\label{Boltzmann2}
\end{eqnarray}
where, $\mathbb{H} = \left(\Phi + R/A + X \left<E_X\right>/m_{\phi}\right)^{1/2}$ is the Hubble expansion rate in terms of new variables.
In the above equation we compute all the dynamical changes with respect to the normalized cosmic scale factor during the reheating period, $A \equiv a/a_I$ with $1/a_I \equiv m_{\phi}$ as an arbitrary scale which is identified with the mass of the inflation. The constants $c_1$ and $c_2$ are defined as
\begin{eqnarray}
	c_1 = \sqrt{\frac{\pi^2 g_*}{30}}\left(\frac{T_{\Gamma}}{m_{\phi}}\right)^{2},~~~~~c_2 = \sqrt{\frac{3}{8\pi}} .
\end{eqnarray}
Here, $M_{pl}(=\sqrt{8\pi}M_p)$ is the Planck mass.
The initial conditions for solving the above set of Boltzmann equations are,
\begin{eqnarray} \label{cond}
	\Phi(1) = \frac{3}{8\pi} \frac{M_{pl}^2 H_I^2}{m_{\phi}^4};~~~~~R(1)=X(1)=0,
\end{eqnarray}
where, the initial Hubble expansion rate is expressed as $H_I^2 = (8\pi/3M_{pl}^2)\rho^{end}_{\phi}$.
The set of Boltzmann equations can be  solved for a given inflaton decay constant $\Gamma_{\phi}$ which, for notational convenience, has been parametrized as,  
\begin{equation}
\Gamma_\phi =\sqrt{\frac{4\pi^3 g_*}{45}} \frac{T_{\Gamma}^2}{M_{pl}}  \ ,
\end{equation}
Notice that $T_{\Gamma}$ here is just a parameter related to the decay rate of inflation. Usually, $T_{\Gamma}$ is identified as the reheating temperature by assuming an instantaneous conversion of inflaton energy into radiation at the instant of reheating$[i.e., \text{when}~H(t) = \Gamma_{\phi}]$. We will define temperature during reheating period in terms of radiation energy density as $
T \equiv T_{\rm rad}=\left[ {30}/{\pi^2 g_*(T)}\right]^{1/4} \rho_R^{1/4}  \ .
$
Hence, as we have mentioned in the Introduction, the reheating temperature $T_{re}$  is measured from the radiation temperature $T_{rad}$ at the instant of maximum transfer of inflation energy into radiation when $H(t)=\Gamma_{\phi}$. 

Another, important bit of information we must keep in mind while connecting reheating with CMB is the existence of maximum radiation temperature during the reheating era\cite{Kolb:1990vq,Chung:1998rq,Giudice:2000ex}. The maximum temperature depends upon the reheating temperature as well as the initial condition of reheating. The approximate analytic expression for the maximum temperature can be obtained as s\cite{Chung:1998rq,Giudice:2000ex}$({\rm i.e.,} \text{~when~}H\gg\Gamma_{\phi} )$
\begin{eqnarray}
T_{max} & \equiv &
\left( \frac{3}{8}\right)^{2/5} \left( \frac{40}{\pi^2}\right)^{1/8}
\frac{g_*^{1/8}(T_{\rm re})}{g_*^{1/4}(T_{\rm max})} M_{p}^{1/4} H_I^{1/4}
T_{\rm re}^{1/2} .
\label{Tmax}
\end{eqnarray}
Depending upon the initial value of the Hubble rate, the maximum temperature can be many orders of magnitude higher than the reheating temperature. Hence, for any physically acceptable model, this temperature must be less than the inflationary energy scale at the end of inflation. The significance of this maximum temperature is that when producing a particle of mass greater than the reheating temperature, the abundance will not be exponentially suppressed by the reheating temperature\cite{Chung:1998rq}.
\subsection{Dark matter relic abundance}
As we have emphasized, our final aim is to study the constraints on dark matter phenomenology through CMB anisotropy. 
Therefore, two essential parameters of our interest would be the current dark matter relic abundance $\Omega_X$, and the CMB scalar spectral index $n_s$. Conventionally the dark matter abundance is expressed in terms of radiation abundance $\Omega_{\rm R}$ $(\Omega_R h^2=4.3\times10^{-5)}$, as
\begin{eqnarray}
	\Omega_X h^2 &=& \frac{\rho_X(T_F)}{\rho_R(T_F)} \frac{T_F}{T_{\rm now}} \Omega_R h^2,\\
	&=& \left<E_X\right> \frac{X(T_F)}{R(T_F)} \frac{T_F}{T_{\rm now}} \frac{A_F}{m_{\phi}} \Omega_R h^2	.
\end{eqnarray}
where $T_F$ is the temperature at a very late time when the universe became radiation dominated and the dark matter, as well as radiation comoving density became constant. The current CMB temperature is given by $T_{\rm now} = 2.35\times10^{-13}{\rm GeV}$. A semianalytic expression for the relic abundance can be arrived at by considering different production mechanisms in different regimes of the thermal evolution. The expressions and their derivation can be found in\cite{Giudice:2000ex}(see also\cite{Allahverdi:2002nb,Allahverdi:2002pu} for an alternative derivation). In the next section, we will see how the dark matter parameter space $(M_X, \langle \sigma v\rangle)$ can be constrained by the CMB anisotropies through the inflationary power spectrum $n_s$. We will consider different inflationary models and their CMB constraints as our input parameters to understand the dark matter phenomenology.  

\section{Constraints from CMB: dark matter phenomenology}
In this section, we explicitly show how the CMB anisotropy can shed light on the dark matter sector considering the present value of its abundance. As emphasized before we will not consider any specific model of dark matter. The main ingredient of our analysis will be a specific model of inflation and its perturbative decay to radiation and then radiation to dark matter during the reheating phase. Considering a specific model of dark matter would be interesting to analyze. However, an important point one should remember when constructing a particle physics model is that all our analyses are at an energy of the order of inflation scale. Therefore, proper high energy modification should be taken into account for any particle physics model of dark matter. Anyway, for the present purpose, we will consider the simplest case as described before. In the subsequent subsection, we first try to illustrate the general procedure to compute the dark matter abundance in terms of the CMB parameter for a chaotic inflation and then we will apply for other models and discuss the constraints. 
 
\subsection{Connecting CMB and reheating via inflation} 
In this section, we will discuss in detail the deep connection between the reheating phase and the CMB \cite{Dai:2014jja}. During inflation, the perturbation modes that became comparable to the horizon are the ones that we observe today. The PLANCK set the pivot scale $k=0.05{\rm Mpc}^{-1}$ for determining the spectral index $n_s$. The comoving Hubble scales $a_kH_k = k$ at $(A)$ and $(D)$ in Fig.{\ref{comoving}
	\begin{figure}[t!]
		\centering	
		\includegraphics[scale=1.5]{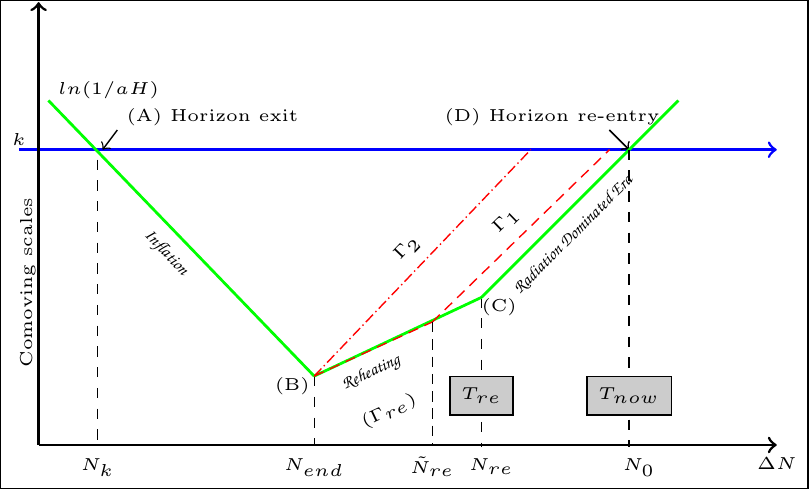}
		\caption{\scriptsize The comoving scales connect the inflationary phase with the CMB. The solution of Boltzmann equations for inflaton decay will connect the end of inflation denoted by the point $B$ and beginning of radiation domination denoted by the point $C$, and the consistent solution exists only for a specific inflaton decay constant $\Gamma_{re}$. All other decay constants $(\Gamma_1,\Gamma_2)$ shown as red lines will not give the correct CMB temperature. Given a particular inflation model, the Boltzmann equations are solve considering three unknown parameters $(\Gamma_{\phi}, \langle \sigma v\rangle, M_X)$. However, imposing two other constraint [Eqs.(\ref{bound})] for our physical universe, we  uniquely fix the value of  $(\Gamma_{\phi}=\Gamma_{re},\langle \sigma v\rangle)$ and consequently the reheating parameters $(N_{re}, T_{re})$ for a given dark matter mass $M_X$. One of the aforementioned constraint equations  essentially sets the correct initial condition for the radiation domination at (C) which evolves to the currently observed CMB through standard big-bang evolution.  In the conventional approach the expansion of the universe during the reheating phase is parametrized by a time-independent effective equation of state $w_{re}$. Therefore, decay of inflaton cannot be directly constrained. Here, however, we have considered the dynamical situation.}
		\label{comoving}
	\end{figure}
	are connected through the reheating period through the following equation
\begin{align}
\label{akhk}
{\rm ln}\left( \frac{a_k H_k}{a_0 H_0}\right) =& -N_k - N_{re} - {\rm ln} \left( \frac{a_{re} H_k}{a_0 H_0}\right).
\end{align}
In order to proceed further specifically from the radiation dominated era to the present CMB time, one important assumption we need to make is that there is no extra entropy production in primordial plasma after reheating. More specifically the entropy is conserved. This assumption is necessary if we want to compute the reheating temperature from CMB: otherwise, we will only be able to give a bound on the reheating temperature through CMB. With this assumption that the reheating entropy is preserved in the CMB and the neutrino background one can arrive at the following relation,
\begin{eqnarray} \label{TreT0}
a(t)^3 s = {\rm const}\implies g_{re}T_{re}^3 = \left( \frac{a_0}{a_{re}}\right)^3 \left( 2T_0^3 + 6\times \frac{7}{8} T^3_{\nu0}\right). 
\end{eqnarray}
where, $s$ is the entropy density. $T_0 = 2.725{\rm K}$ is the present CMB temperature, and  $T_{\nu0}= (4/11)^{1/3}T_0$ is the neutrino temperature and $g_{re}$ is the effective number of light species. $H_0$ is the present value of the Hubble parameter. Therefore, combining the above two equations, one arrives at the following important equation, 
\begin{equation}
T_{re} = \left(\frac{43}{11 g_{re}}\right)^{\frac{1}{3}}  \left(\frac{a_0 T_0}{k} \right) H_k e^{-N_k} e^{-N_{re}} .
\label{TreEq}
\end{equation}
This equation thus establishes the connection between the CMB anisotropy with the reheating temperature once we know the e-folding number during reheating $N_{re}$.

Now, we can have two ways to determine $N_{re}$: (i), solve the scale factor and the evolution equation for the total energy density during reheating by using an effective equation of state parameter($w_{re}$)  \cite{Dai:2014jja} of the fluid comprising inflaton and radiation during reheating, or, (ii) explicitly solve the Boltzmann equation for decaying inflaton during reheating. 
	The first method has been widely studied in the literature. For the convenience of the reader, let us note down the expression of  $N_{re}$ in terms of inflationary observables and reheating parameters following the references \cite{Dai:2014jja,Drewes:2017fmn},
	\begin{equation}
	 N_{re} = \frac{4}{3w_{re}-1}\left[N_k + {\rm ln}\left(\frac{k}{a_0 T_0}\right) + \frac{1}{4}{\rm ln}\left(\frac{40}{\pi^2 g_{\ast}}\right)+ \frac{1}{3}{\rm ln}\left(\frac{11g_{\ast}}{43}\right) - \frac{1}{2}{\rm ln}\left(\frac{\pi^2M_p^2~rA_s}{2V^{\frac{1}{2}}_{\rm end}}\right) \right] .
	 \label{nre_dkw}
	\end{equation}
where $N_k,~r,~A_s$, etc. are known for specific inflationary models in terms of the spectral index $n_s$. $w_{re}$ is assumed be to an effective  time-independent equation of state during reheating. The main disadvantage of this method is that it does not shed light on the microphysics of the reheating phase and its 
	effect on the subsequent evaluation. We propose the second method \cite{Maity:2017thw} with added advantages that we have largely exploited in this paper. We have also stated the limitations of our approach and possible extensions. For both the cases the initial conditions will be at point "B" in Fig.\ref{comoving} which is set by slow roll inflation constrained by the CMB observation. This connection is clearly depicted in Fig.(\ref{comoving}). From the figure, it is clear that a particular inflationary model with a scalar spectral index $n_s$ sets unique initial conditions for the Boltzmann equations for decaying inflaton and its decay products during reheating. And in this phase, one of the important parameters is the inflaton decay constant $\Gamma_{\phi}$ that controls the dynamics with a strong constrain that the dominant energy component will be the inflaton and the radiation. This requirement fixes a specific value of $\Gamma_{\phi}=\Gamma_{re}$ for which Boltzmann equations predict a particular reheating e-folding number $N_{re}$ and reheating temperature $T_{re}$ which finally evolves to the current value of the CMB temperature $T_0 = 2.7{\rm K}$. Hence, the first part of our calculation is to figure out $\Gamma_{re}$, Finally solving the Boltzmann equations has added advantages as opposed to the conventional effective equation of state method. As mentioned earlier, because of considering explicit decay of inflaton, apart from radiation we can easily consider an addition component such as dark matter in our analysis. Because of the constraint of dark matter abundance in the present universe, we can establish a direct connection between the CMB anisotropy and the dark matter phenomenology. Therefore, this approach will lead us to establish a direct connection between the CMB and the dark matter through the inflation and reheating.

\subsection{Methodology: CMB to dark matter via reheating}
Let us now summarize again the connection between the CMB and dark matter phenomenology via reheating. The CMB power spectrum provides the initial conditions for the reheating phase through inflationary observables. While the CMB temperature is intimately connected to the reheating temperature, the reheating phase links the end of inflation and the beginning of the radiation phase parametrized by the reheating temperature. All the particles including dark matter in the universe were created during the phase of reheating through inflaton decay. Therefore, we can clearly understand the deep connection between the CMB and the dark matter we see today via the reheating phase.
In this section, we will discuss the methodology toward establishing this connection between the CMB anisotropy and the dark matter phenomenology we just mentioned.
For any general canonical inflation model, we first identify the inflation model dependent input parameters such as $[N_k, H_k, V_{\rm end}(\phi_k)]$ for a particular CMB scale $k$ (CMB pivot scale) at its horizon crossing. As has been pointed out before, given a canonical inflaton potential $V(\phi)$, the inflationary e-folding number $N_k$  and Hubble constant $H_k$ can be expressed as  
\begin{equation}
H_k = \frac{\pi M_p \sqrt{r A_s}}{\sqrt{2}}~~
;~~	N_k = {\rm ln}\left(\frac{a_{\rm end}}{a_k} \right) = \int_{\phi_k}^{\phi_{\rm end}} \frac{H}{\dot{\phi}}d\phi =  \int_{\phi_k}^{\phi_{\rm end}} \frac{1}{\sqrt{2\epsilon_V}} \frac{|d\phi|}{M_p} .
\label{Hk}
\end{equation}
In order to define the above quantities, we use the following slow-roll approximated equations  
\begin{equation}
3H\dot{\phi}= -V'(\phi)~~;~~H_k^2 = \frac{V(\phi_k)}{3M_p^2}.
\label{HkSR}
\end{equation}
where the field value $\phi_{\rm end}$ is computed form the condition of the end of inflation,
\bea
\epsilon(\phi_{end}) = \frac{1}{2 M_p^2} \left(\frac{V'(\phi_{\rm end})}{V(\phi_{\rm end})}\right)^2 =1,
\eea
while, the field $\phi_k$ at the horizon crossing in terms of the scalar spectral index $n_s^k$ can be found by inverting the following equation:
\begin{equation}
n^k_s = 1 - 6\epsilon(\phi_k) + 2\eta(\phi_k).
\end{equation}
Once we identify all the required parameters from the inflation, the subsequent reheating phase will be described by the appropriate Boltzmann equations (\ref{Boltzmann2}) and also the background dynamics for the scale factor $a$. As emphasized earlier we will consider all the decay process to be perturbative. During the reheating phase, one of the important parameters is the reheating e-folding number $N_{re}$. It connects the scale factor between the end of inflation $a_{end}$ and the end of reheating $a_{re}$ with the following definition $N_{re} = {\rm ln}(a_{\rm re}/a_{\rm end})$. In order to establish the relation among the reheating temperature $T_{re}$, the inflationary index $n_s$, and dark matter parameters $(M_X,\sigma)$ we simultaneously solve the set of Boltzmann equations (\ref{Boltzmann2}) with the following three initial conditions for three components of energy density, 
\bea
\Phi(1) = \frac{3}{8\pi} \frac{M_{pl}^2 H_I(n^k_s)^2}{m_{\phi}^4};~~~~~R(1)=X(1)=0 .
\label{reheating}
\eea
While solving Boltzmann equations we simultaneously satisfy the following two constraint equations
\bea
\Omega_X h^2 = 0.12 ~~;~~T_{re} = \left(\frac{43}{11 g_{re}}\right)^{\frac{1}{3}}  \left(\frac{a_0 T_0}{k} \right) H_k e^{-N_k} e^{-N_{re}} ,
\label{bound}
\eea
which are related to current dark matter abundance, and evolution of $T_{re}$ to current CMB temperature $T_{0} = 2.7 {\rm K}$. 
Therefore, we essentially solve the Boltzmann equations (\ref{Boltzmann2}) starting from the end of inflation till the dark matter freezes out considering constraints equations (\ref{bound}).

Once the dark matter freezes out to the current value of  dark matter abundance, one of the dark matter parameters, for instance, the cross section $\langle \sigma v\rangle$ can be fixed for a given set of values of $(\Gamma_{\phi}, M_X)$. By using further condition on the end of reheating with the e-folding number $N_{re}= {\rm ln}(a_{re}/a_{\rm end})$,
\bea
H(a_{re})^2= \frac{\dot{a}_{re}}{a_{re}} =\frac{8\pi}{3M_{Pl}^2} \left[\rho_\phi(\Gamma_{\phi},M_X) +\rho_R(\Gamma_{\phi},M_X) +\rho_X(\Gamma_{\phi},M_X) \right] = \Gamma_{\phi}^2 ,
\label{recond}
\eea
we fix the value of $\Gamma_{\phi}$ in terms of scalar spectral index $n_s$ and the dark matter mass $M_X$.
In the above expression all the energy densities are written as a function of $(\Gamma_{\phi}, M_X)$ at the end of reheating. Upon getting the solution for all the energy components we express the reheating temperature  as  
\bea 
T_{re} \equiv T^{end}_{rad}=\left[ {30}/{\pi^2 g_*(T)}\right]^{1/4} \rho_R(\Gamma_{\phi},n_s,M_X)^{1/4}.
\eea 
where, radiation energy density is computed at the end of reheating 
\begin{eqnarray}
\rho_R(\Gamma_{\phi},n_s,M_X)= \frac {R~ m_{\phi}^4}{A^4} \Big|_{H=\Gamma_{\phi}}
\end{eqnarray}
This is the temperature of the radiation component at the end of reheating. In our numerical analysis, we will feed this definition of reheating temperature into the Eq.(\ref{bound}).  Hence for a given dark matter mass, the reheating temperature will be fixed by the spectral index $n_s$. As mentioned earlier the connection between the reheating temperature and the inflation scalar spectral index was first pointed out in \cite{Dai:2014jja}. After solving all the above equations we are left with one free parameter that is the mass of the dark matter $M_X$.

 With this strategy in hand, we will numerically solve the  Boltzmann equations starting from the end of inflation and show how for a specific dark matter mass $M_X$ one can constrain the dark matter annihilation cross section through the CMB anisotropy. For this, we will consider some specific inflationary models. As we have mentioned, we will use the CMB pivot scale of PLANCK, $k/a_0 = 0.05 {\rm Mpc}^{-1}$.
All the quantities of our interest such as $(T_{re},N_{re}, \langle \sigma v\rangle)$ will be studied at  the aforementioned scale with respect to the inflationary power spectrum $n_s =0.9659 \pm 0.0082$ for $\Lambda CDM+r$ model from Planck data.

At this point we must mention that the production of dark matter prior to the nucleosynthesis era may have important consequences on the subhorizon perturbations of the radiation and the dark matter\cite{Erickcek:2011us} and may also affect the annihilation rate of the dark matter\cite{Erickcek:2015jza}. A detailed study of these effects is done by following the evolution equations for perturbations of the above three components and the appropriate transfer function. These studies are beyond the scope of the present work and will be considered in a future publication.

\subsection{Chaotic inflation: General results}

To elucidate our method and discuss the general results, in this section we discuss the chaotic inflationary model in detail. For all the other models we will see the qualitative behavior will be the same. The chaotic type models are represented by the power-law potentials of the form:
\begin{equation}
	V(\phi) = \frac{1}{2} m^{4-n} \phi^n .
\end{equation}
where $m$ is the mass scale associated with the inflation. The initial conditions for Boltzmann equations are provided by the inflation energy density at the beginning of the reheating, which in turn will depend on the inflationary power spectrum $n_s$. To establish such a connection, and its effect on the subsequent evolution we compute the field value at the end of inflation $\phi_{\rm end} = M_p \frac{n}{\sqrt{2}} $ using the condition for the end of slow roll inflation $\epsilon(\phi) =1$. Therefore, using this we get the initial condition for the reheating phase as defined in Eq. (\ref{cond}) 
\begin{eqnarray} \label{bc}
\Phi(1) = \frac{3}{8\pi} \frac{M_{pl}^2 H_I^2}{m_{\phi}^4} \simeq \frac{4 V_{end}}{3 m_{\phi}^4} = \frac{2}{3} \frac{m^{4-n}}{m^4_{\phi}} \left( \frac{n M_p}{\sqrt{2}} \right)^n ;~~~R(1)=X(1)=0.
\end{eqnarray}
Other important quantities that are directly connected with the CMB anisotropy through the relations equation (\ref{akhk}) are
\begin{equation}
H_k = \frac{\pi M_p \sqrt{r_k A_s}}{\sqrt{2}} =  \frac{\pi M_p \sqrt{\frac{8n}{n+2}(1-n^k_s)
		 A_s}}{\sqrt{2}}
~~~;~~	N_k = {\rm ln}\left(\frac{a_{\rm end}}{a_k} \right) =   \left[ \frac{n+2}{2(1-n^k_s)} -  \frac{n}{4} \right] ,
\label{hknk}
\end{equation}
where, the scalar spectral index  $n^k_s$ and consequently the tensor to scalar ratio $r_k$, for a particular CMB scale $k$ are expressed in terms of the inflaton field as 
\begin{equation}
 n_s^k = 1 - \frac{2n(1-n)M_p^2}{\phi_k^2} - \frac{3n^2 M_p^2}{\phi_k^2} ~~;~~r_k =\frac{8n}{n+2}(1-n^k_s) .
  \label{rns}
\end{equation}
$\phi_k$ is the inflaton field value for a particular scale $k$. 
And finally, using Eqs. (\ref{Hk}),~(\ref{HkSR}),~(\ref{hknk}) and (\ref{rns})the parameter $m$ in terms of the spectral index is found to be
\begin{equation}
 m = M_p (3\pi^2 r A_s)^{\frac{1}{4-n}} \left[ \frac{1-n_s}{n(n+2)}\right]^{\frac{n}{2(4-n)}} .
\end{equation}
Another important quantity before solving the Boltzmann equations is to know the equation of state parameter, which for the power-law potential is given in Eq.(\ref{eos}). For $n=2$, the homogeneous inflaton field will behave as pressure-less dust with equation of state $w_{\phi}=0$. 

Now, in order to establish the relation between the reheating temperature $T_{re}$ and the inflationary index $n_s$, we follow the methodology explained before. The numerical procedure would be to first solve the set of Boltzmann equations (\ref{Boltzmann2}) considering inflaton decay constant, $\Gamma_{\phi}$ and annihilation cross section $\left< \sigma v \right>$ as free parameters. The initial condition for the inflaton energy density is fixed by the spectral index as discussed earlier. Once the solution for the radiation energy density during reheating is known, we simultaneously solve Eqs.(\ref{recond}) and (\ref{TreEq}) relating the reheating temperature with the current CMB temperature in a self-consistent manner.

For any other model, we will follow the same procedure discussed above. As we have already mentioned and elaborated in the Introduction, in the usual reheating constraint analysis \cite{Dai:2014jja}, the connection between the inflationary parameters $(n_s^k, N_k)$, the reheating parameters $(T_{re}, N_{re})$ and the CMB temperature $T_0$ are established based on two important assumptions. First one is the effective single fluid description of the reheating phase with a time independent equation of state. The second assumption is that the inflaton energy is completely transferred into radiation at an instant  $H=\Gamma_{\phi}$. We have already stressed earlier that those two assumptions are obviously not correct. In addition, we also have considered an additional dark matter field in the picture. Therefore, we compare our result with the usual formalism and the difference will be displayed in various plots.   
 \begin{figure}[t!]
 	\centering	
 	\subfigure[]{\includegraphics[scale=0.45]{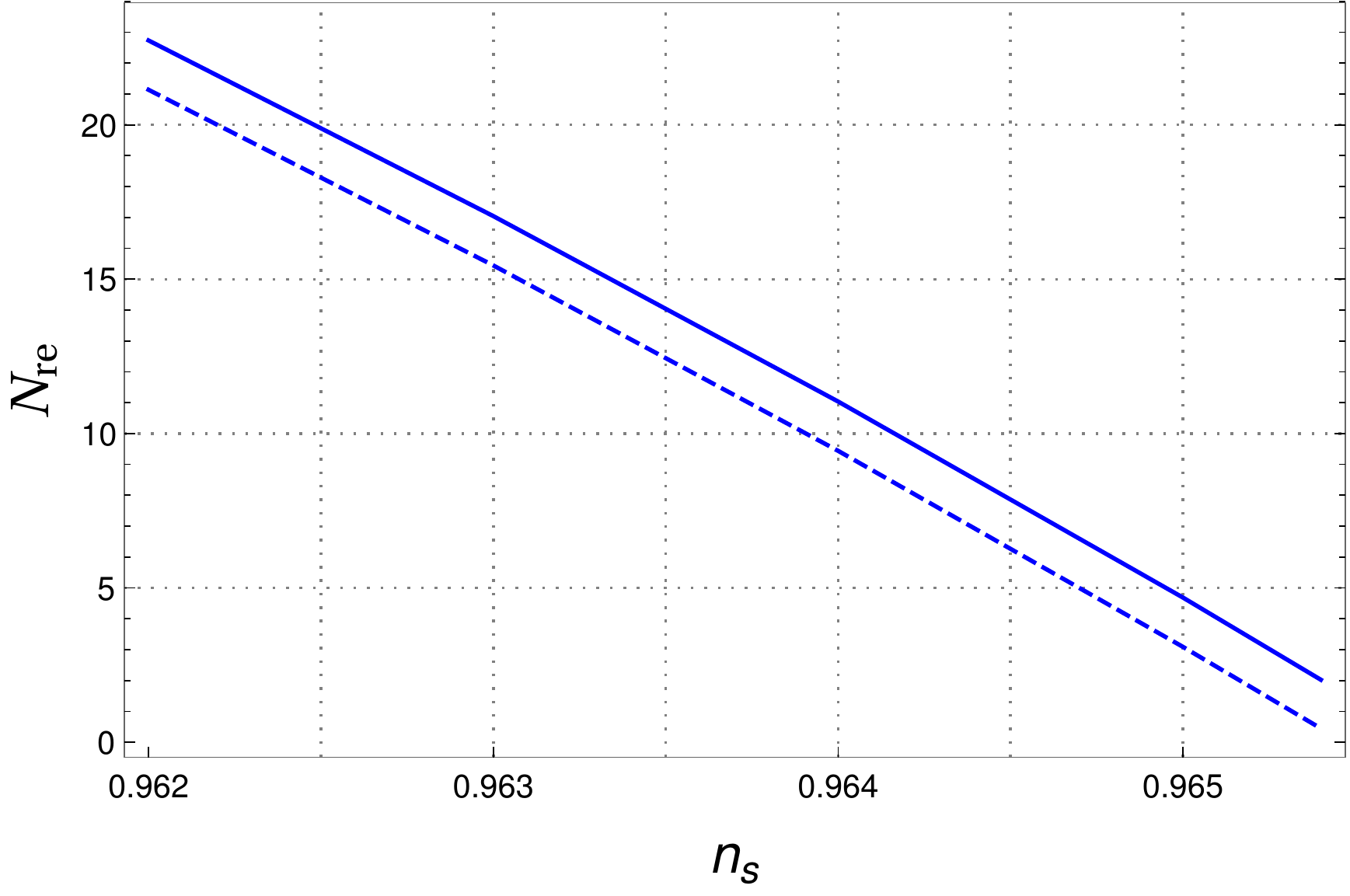}}
 	\subfigure[]{\includegraphics[scale=0.452]{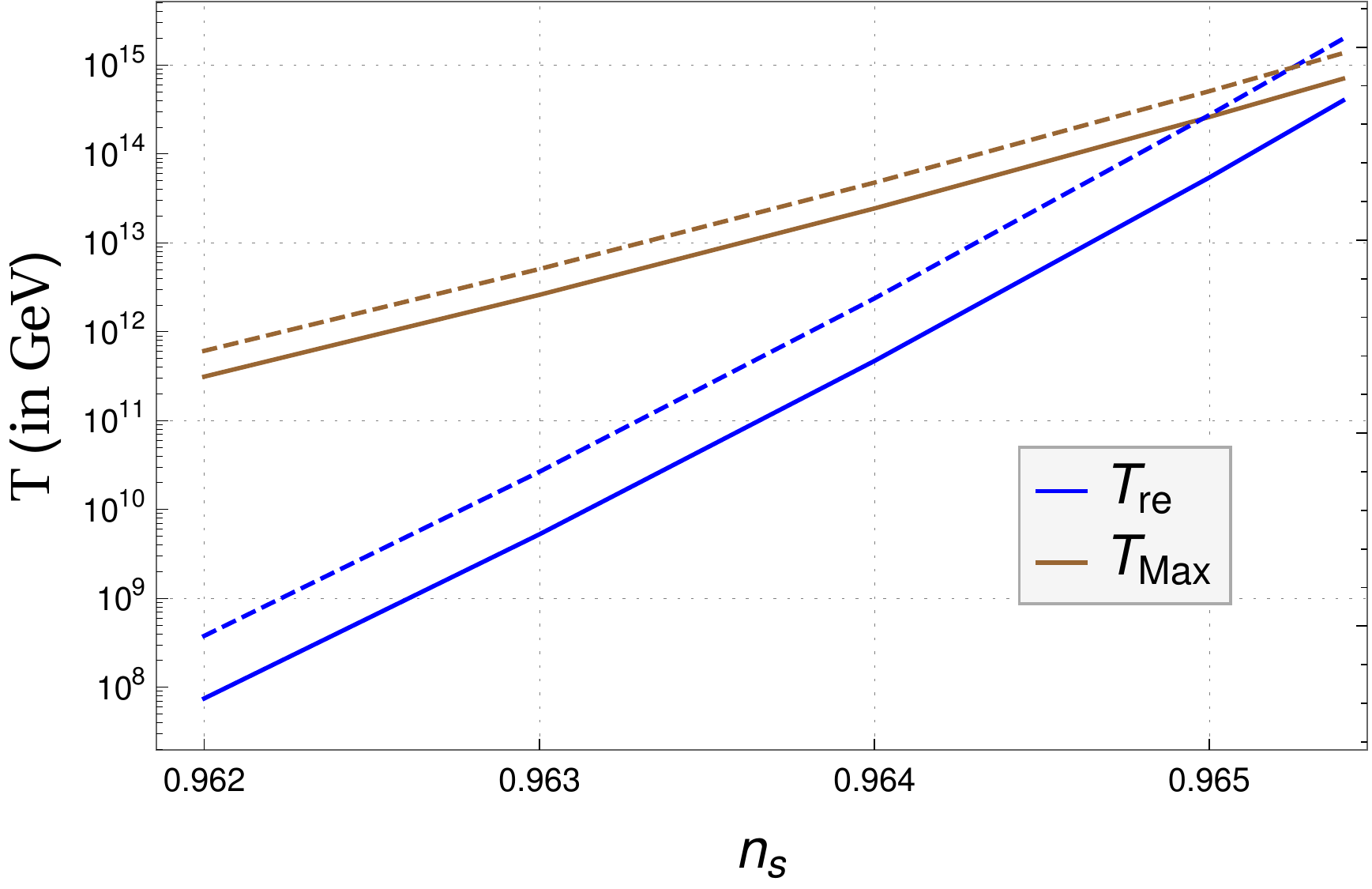}}
 	\caption{\scriptsize Variation of (a) reheating e-folding number $N_{re}$ and (b)the reheating temperature $T_{re}$ and the maximum radiation temperature $T_{\rm max}$ with respect $n_s$ have been plotted. For comparison, dashed lines are shown from the Ref. \cite{Dai:2014jja} where the complete conversion from inflaton to radiation has been assumed. We clearly see the order of magnitude difference in the temperature at the moment we include the explicit decay of inflaton in the reheating analysis \cite{Maity:2017thw}. These two plots are independent of dark matter masses for a set of given initial conditions.}
 	\label{ch_nsnret}
 \end{figure}
\begin{figure}[t!]
	\centering	
	\subfigure[]{\includegraphics[scale=0.45]{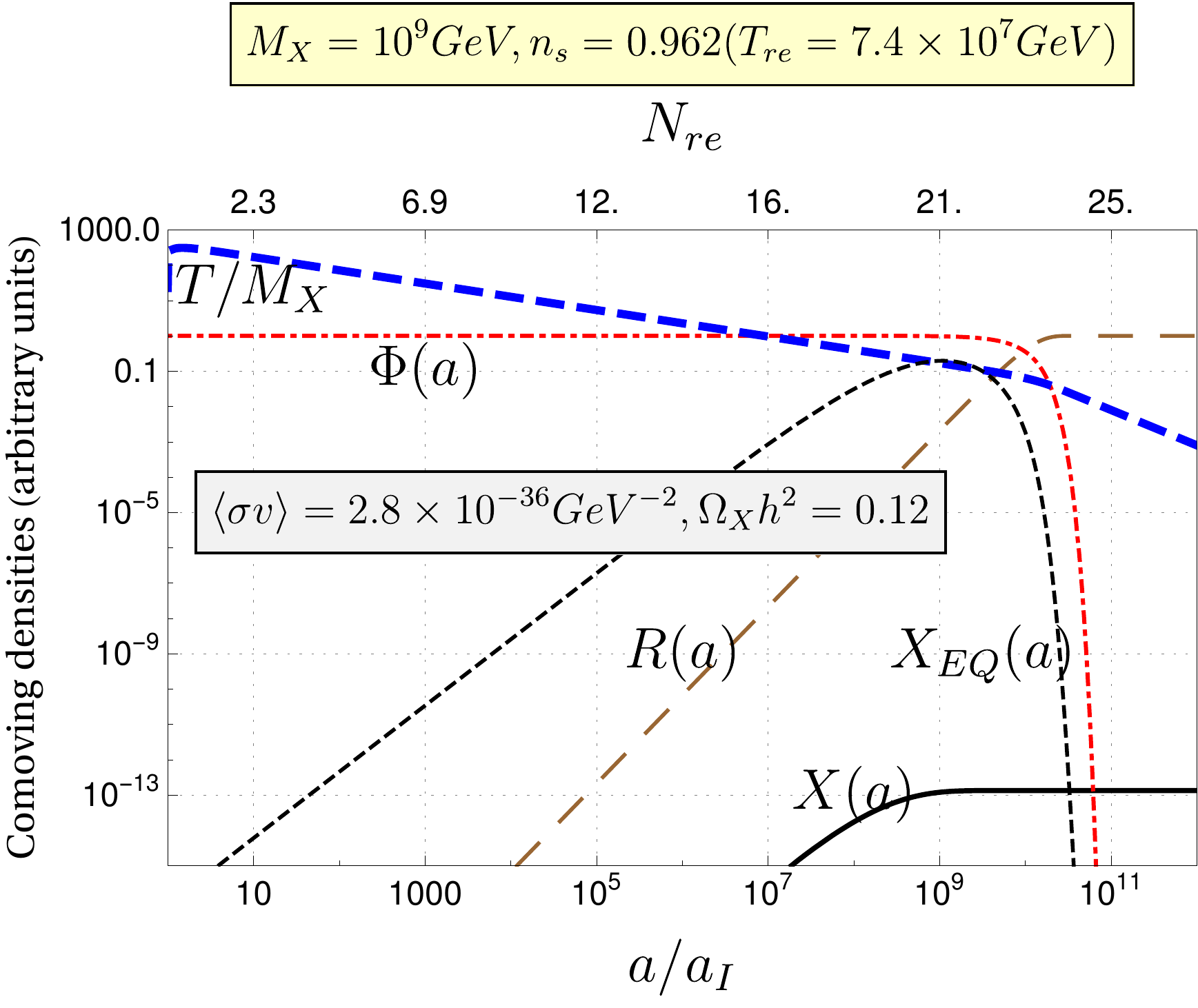}}
	\subfigure[]{\includegraphics[scale=0.452]{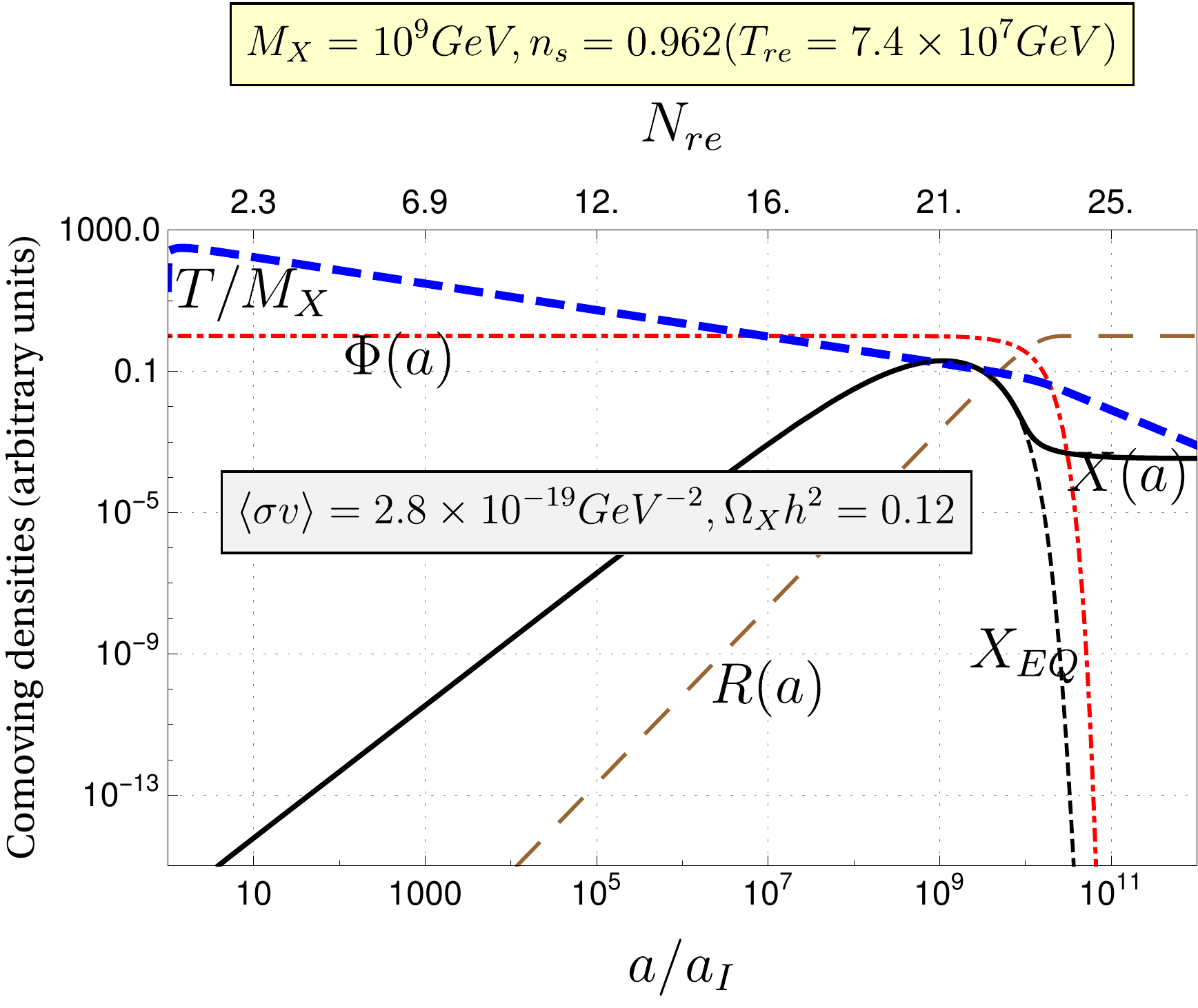}}
	\caption{\scriptsize We have two different dark matter production mechanisms as described in the text: (a) Freeze-in and (b) Freeze-out as discussed. Choosing the same dark matter mass and reheating temperature, we can realize these two production mechanisms depending upon its annihilation cross section $\langle \sigma v\rangle$. The figures here show the evolution of different components(in some suitable units) the inflaton(red dot dashed line), the radiation(brown dashed line) and the temperature(thick blue dashed line) with the normalized scale factor( alternatively, the e-folding number after the end of inflation). Black dashed lines show the evolution of equilibrium dark matter distribution while the black solid line is for the dark matter. In this work we will exclusively assume the dark matter production via freeze-in mechanism when connecting the current relic abundance with CMB.}
	\label{inout}
\end{figure}

Including the dark matter component in the reheating constraint analysis and generalizing the formalism given in \cite{Maity:2017thw}, 
we will solve the system of Boltzmann equation (\ref{Boltzmann2}) taking inflaton decay rate $\Gamma_{\phi}$ as a free parameter. For this, the initial condition is set by the CMB power spectrum via inflation as given in Eq.\ref{bc}. From our analysis, we will see that one of the free parameters $\Gamma_{\phi}$ will be fixed by $n_s$ through reheating temperature [see (\ref{TreEq})]. At this point, there are several important questions we will ask such as a) Does the dark matter mass have any effect on the reheating temperature? As we have already stated in the Introduction, b) Does the CMB play any role in understanding the properties of dark matter and its production mechanisms?

Throughout the subsequent discussions, we will try to answer the aforementioned questions. Even though the dark matter will play an important role after reheating we have not found any significant effect of it's mass or the annihilation cross section on $(T_{re},N_{re})$ provided the produced dark matter relic abundance is within the current dark matter relic abundance. In Fig.\ref{ch_nsnret}, we have plotted $(N_{re}, T_{re})$ with respect to $n_s$. An important observation is the existence of a maximum reheating temperature where two radiation temperatures  $T_{\rm max}$ and $T_{re}$ meet at around $(n^{\rm max}_s \simeq 0.9656,  T_{re}^{\rm max}\simeq 10^{15} $GeV). This is the point where the reheating process is almost instantaneous. If we consider $1 \sigma$ range of $n_s$ from PLANCK, one also gets minimum reheating temperature $T^{\rm min}_{re}\simeq 6\times 10^{7}$ GeV for $n_s \simeq 0.962$. At this point let us emphasize the difference between the result of our analysis(solid line) and the usual reheating constraint analysis(dashed line) following the Ref. \cite{Dai:2014jja}. It clearly shows one order of magnitude difference in reheating temperature. The source of this difference is coming from the incomplete decay of inflaton to radiation field. Finally, we numerically fit the data, and the relation between the reheating temperature $T_{re}$ and spectral index $n_s$ is found as, 
\begin{equation}
	{\rm log}\left( T_{re}\right) \simeq Q_p\left[A + B(n_s - 0.962) + C(n_s - 0.962)^2 \right]. 
	\label{nsTre}
\end{equation}
where, the dimensionless constants $A=8$, $B =1.8\times10^3$ and $C=5.5\times 10^4$ turned out to be almost model independent. The reason may have its origin in the same mechanism that is responsible for the inflaton decay into the radiation. Model dependence in the above expression for reheating temperature comes only through the parameter $Q_p$. To complete the discussion, let us mention here that for a chaotic and $\alpha$-attractor model with $\alpha=1$, the value of $Q_p$ turned out to be unity. Also, our numerical fitting shows that for different $\alpha$ values, $Q_p \sim {\rm log}_{10}(\alpha)/\alpha^{1/2}$ and for natural inflation $Q_p \propto 1/f_b$.

  \begin{figure}[t!]
  	\centering
  	\subfigure[]{\includegraphics[scale=0.45]{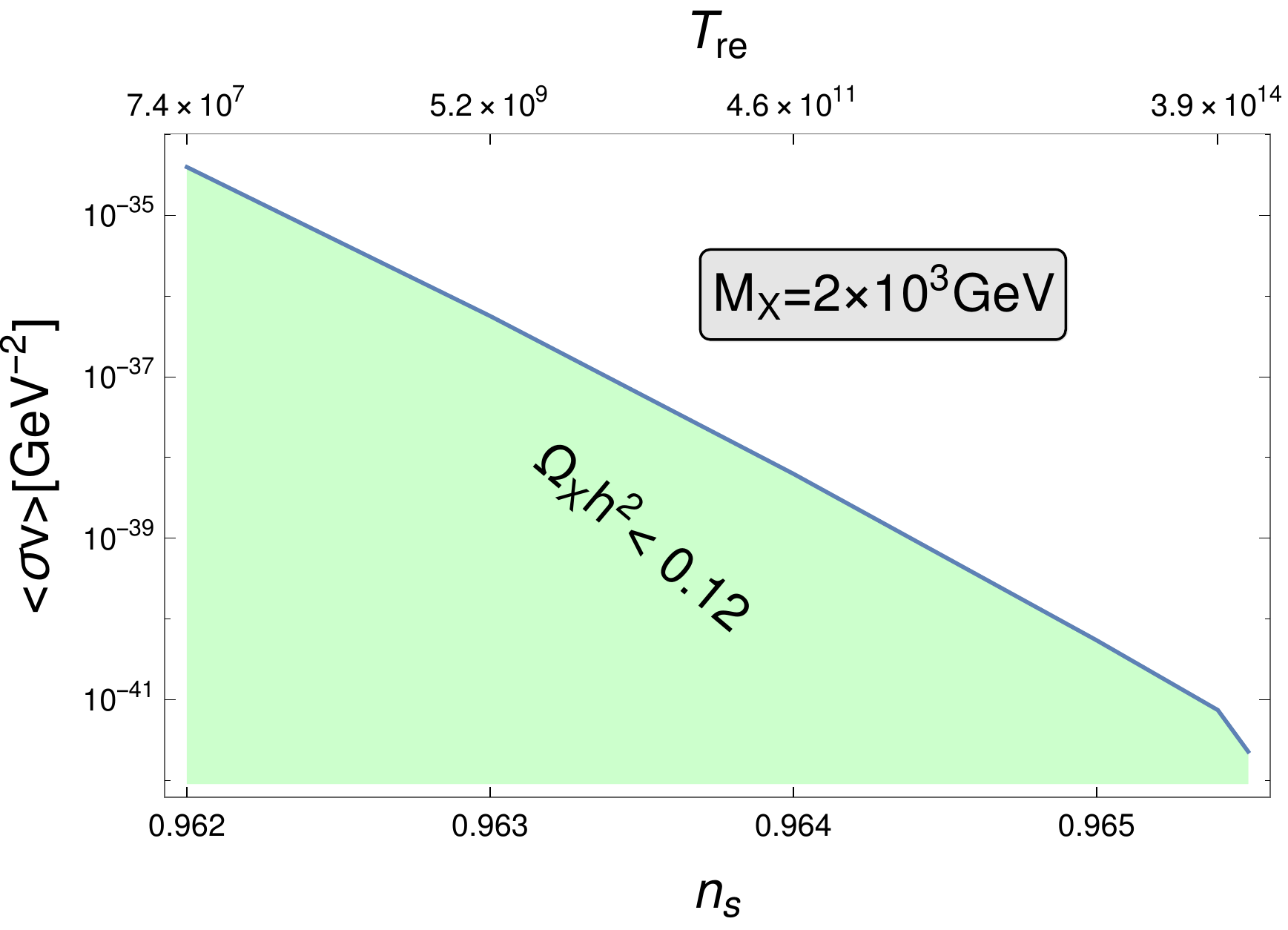}}
  	\subfigure[]{\includegraphics[scale=0.454]{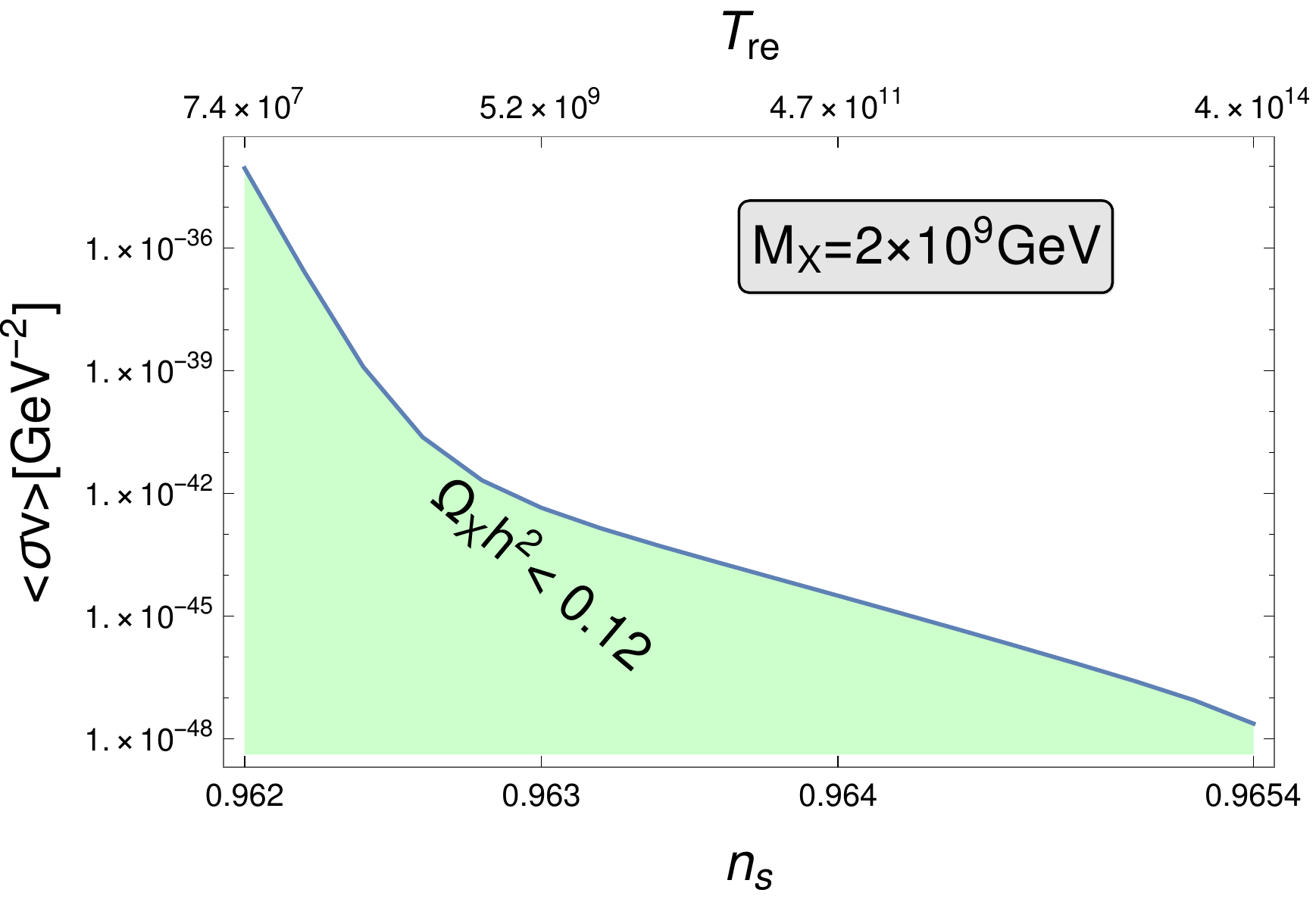}}
  	\caption{\scriptsize For fixed dark matter mass, we have plotted the contour of $\Omega_X h^2 = 0.12 $ in the $n_s$-$\langle \sigma v\rangle$ plane. The reheating temperature that is fixed once we know the spectral index is also plotted on the upper axis. The shaded region below the contour line is the parameter space allowed by current dark matter abundance. Here, we have considered the dark matter masses (a)$10^9{\rm GeV}$ and (b)$10^3{\rm GeV}$ for the chaotic $m^2\phi^2$ model.}
  	\label{ccs}
  \end{figure}
 Let us now turn to the question-b, which is the main purpose of this work. In the previous section, we have established one-to-one correspondence between $n_s$ and $T_{re}$. This fact provides us a way to figure out the direct connection between the CMB anisotropy and the dark matter via reheating. Before we discuss the constraints, we emphasize again the fact that dark matter production mechanism can be either freeze-in or freeze-out depending upon the couplings as has been discussed in the Introduction. However, we will consider the dark matter production via the freeze-in mechanism in this work. However, let us emphasize the fact that for $M_X \gg T_{re}$, freeze-in is the only mechanism that satisfies correct dark matter abundance namely $\Omega_X h^2 \leq 0.12$. This is also clearly seen for a specific case shown in the Fig.\ref{inout}. For $M_X < T_{re}$, we have only considered the dark matter production via the freeze-in mechanism. We will study other mechanisms in more detail in our subsequent publication. Given a specific mechanism, we constrain the dark matter parameter space depending upon a specific inflationary model.
 In Fig.\ref{ccs}, we have plotted annihilation cross section $(\langle \sigma v\rangle, n_s)$ for different dark matter masses considering specific chaotic model $n=2$. The important point one infers from those plots is that the CMB temperature correlation can directly constrain the dark matter parameter space $(M_X,\langle \sigma v\rangle)$ through the inflationary power spectrum $n_s$. For a given value of $n_s$, one can precisely predict the value of annihilation cross-section once the dark matter mass is fixed. As an example given a dark matter mass $M_X=2 \times 10^3$ GeV, CMB anisotropy restricts the annihilation cross section within $10^{-35} {\rm GeV}^{-2} > \langle \sigma v\rangle > 10^{-41} {\rm GeV}^{-2}$ for the $2 \sigma$ region of $n_s$.
 
 Depending upon the value of dark matter mass our main results of the current paper are the following important relations:
  i) If $M_X>T_{re}$, the dark matter freezes in before the reheating and the relic abundance  for a fixed dark matter mass behaves as $\Omega_X h^2 \propto \left< \sigma v \right> T_{re}^7$ \cite{Chung:1998rq,Giudice:2000ex}. Therefore, we established an important relation between the annihilation cross section $\left< \sigma v \right>$ and the scalar spectral index $n_s$ considering the current value of the dark matter relic abundance as
  \begin{equation} \label{sigmans1}
  \left< \sigma v \right>\Big|_{M_X>T_{re}} \propto 10^{-7A -7 B(n_s - 0.962) - 7 C(n_s - 0.962)^2} .
  \end{equation}
  
  ii) In a similar manner, for $M_X < T_{re}$, the dark matter freezes in during the radiation dominated phase following the relation $\Omega_X h^2 \propto \left< \sigma v \right> T_{re}$\cite{Dev:2013yza}. In this case also we will have the following important relation in a different dark matter mass regime: 
  \begin{equation} \label{sigmans2}
  \left< \sigma v \right>\Big|_{M_X<T_{re}} \propto 10^{-A - B(n_s - 0.962) - C(n_s - 0.962)^2} .
  \end{equation}
   
  So far, all our important findings were based on the chaotic inflation. In the subsequent sections we will consider various other prominent inflationary models.   

\subsection{Natural inflation}
The natural inflation model\cite{Freese:1990rb, Freese:2014nla} proposed in the early 1990s is one of the best theoretically motivated models of inflation. The prediction of this model is marginally consistent with the recent observations.\footnote{It has been shown in\cite{Gerbino:2016sgw} that by considering the neutrino properties in calculating $n_s$, this model may comply well with observation.} The inflationary potential in this case is given by

\begin{equation}
V(\phi) = \Lambda^4 \left[ 1 - \cos\left(  \frac{\phi}{f}  \right) \right] .
\end{equation}
  where $\Lambda$ is the height of the potential setting the inflationary energy scale, and $f$ is the width of the potential known as the axion decay constant in particle physics. To be consistent with the CMB data this model needs a super-Planckian value of the axion decay constant. We have taken $f=10M_p$ and $f=50M_p$ for illustration. During the reheating, potential may be approximated as a power-law potential by expanding it around the minimum as long as $\phi < f$
  \begin{equation}
   V(\phi) \simeq\frac{1}{2} \frac{\Lambda^4}{f^2}\phi^2
  \end{equation}
From this expression of the potential it is easy to identify the inflaton mass by tree-level expression
\begin{equation}
 m_{\phi} = \frac{\Lambda^2}{f}
\end{equation}
While the inflation equation of state from Eq.(\ref{eos}) is found to be $w_{\phi}=0$.
\begin{figure}[t!]
	\centering
	\subfigure[]{\includegraphics[scale=0.45]{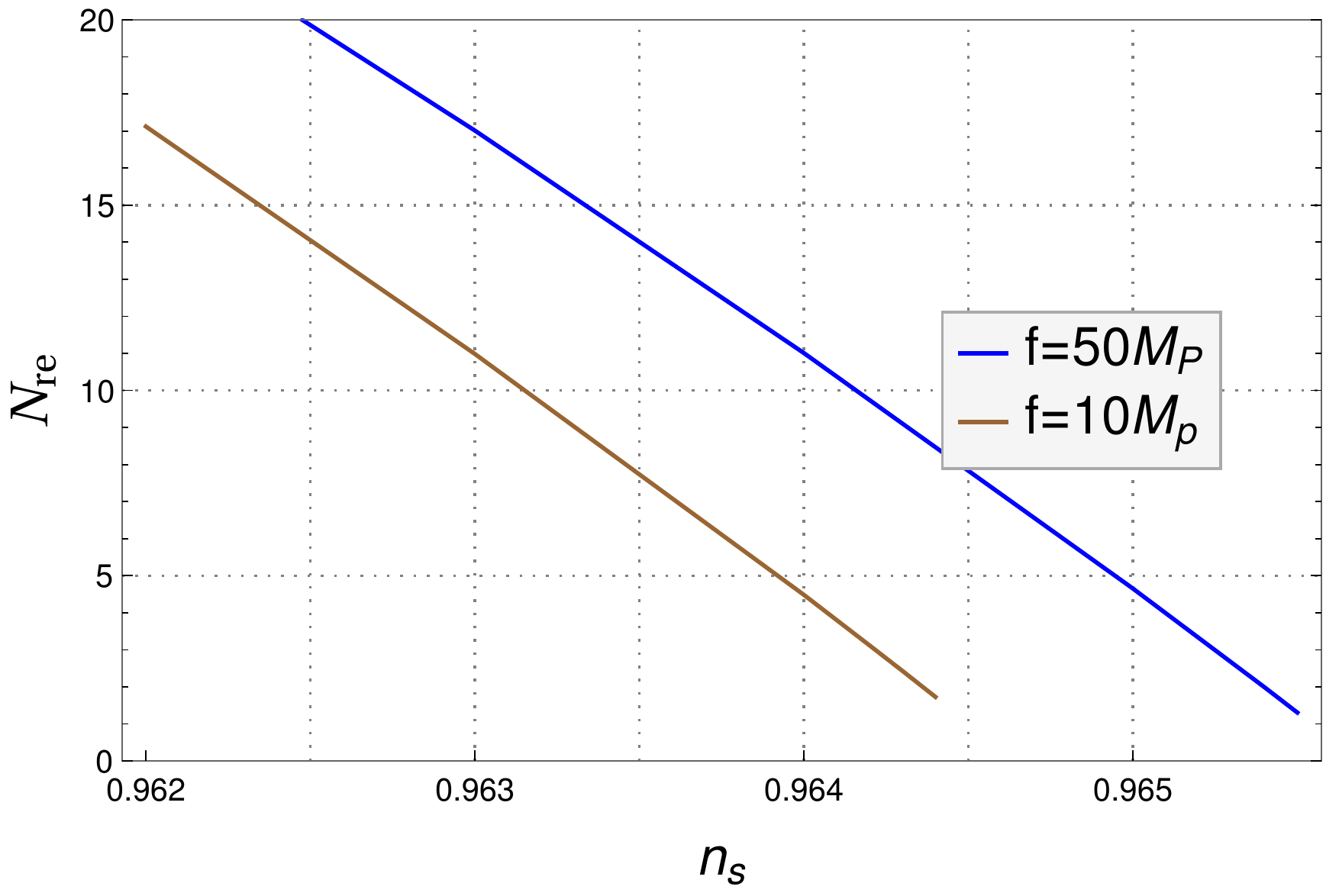}}
	\subfigure[]{\includegraphics[scale=0.45]{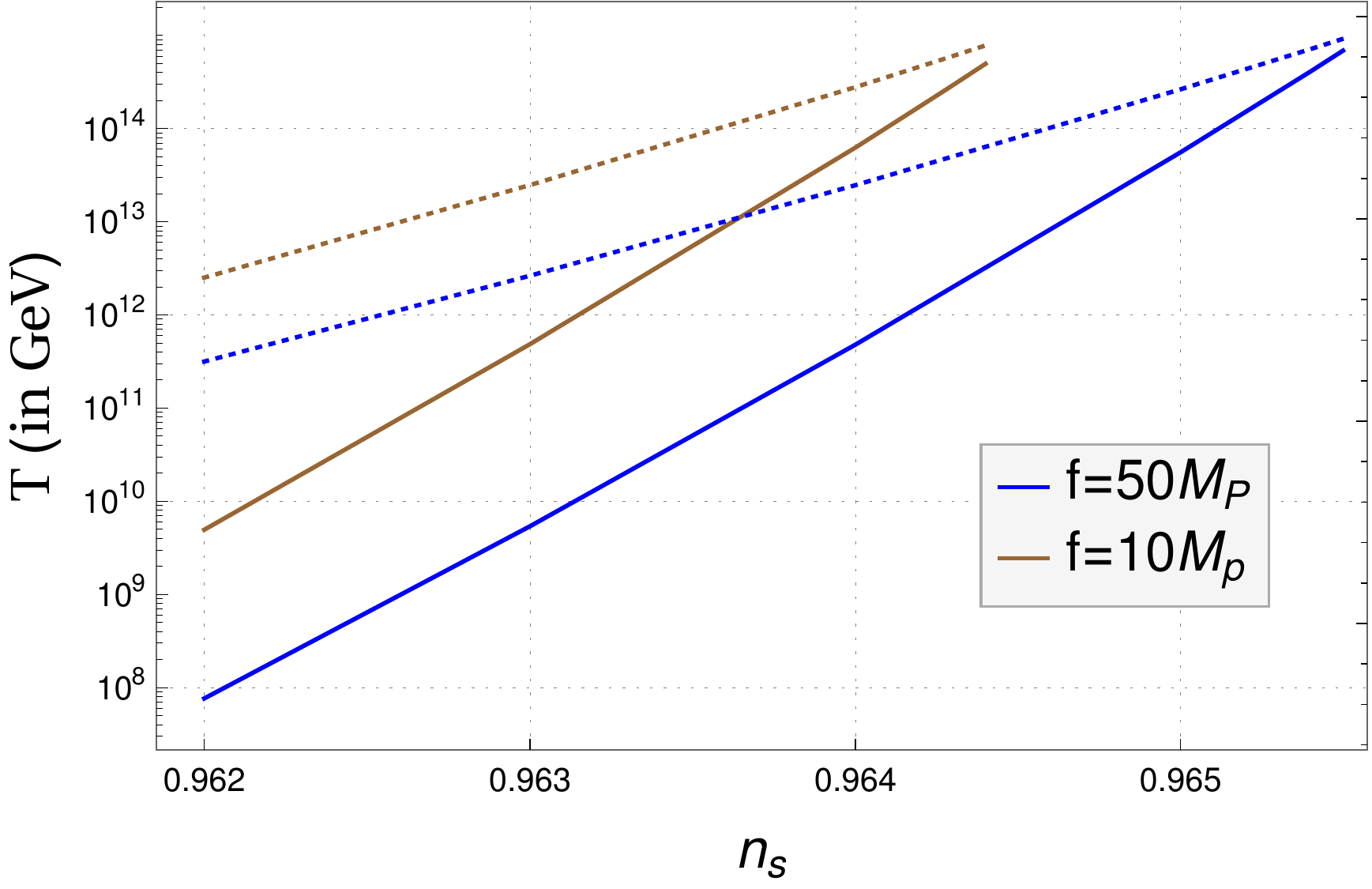}}
	\caption{Variations of (a) $N_{re}$  and (b) $(T_{re}, T_{\rm max})$ with respect in $n_s$ have been plotted for for axion decay constant $f=(10, 50){\rm M_p}$. The duration of reheating increases with $f$, and as a result the reheating temperature decreases with increasing $f$.}
	\label{ax_nsnret}
\end{figure}

The CMB normalization defined as $A_s$ fixes the value of  $\Lambda \simeq 10^{16}$ GeV. Therefore, by tuning the value of the axion decay constant $f$ we can fit model with respect to the observation. For the usual quadratic axion potential near its minimum, we consider effective equation state $w = 0$ during reheating. From Fig.\ref{ax_nsnret} the behavior of the $(N_{\rm re}, T_{\rm rad})$ in terms of $n_s$ can be summarized as follows: with decreasing $f$, the model becomes increasingly disfavored as it is going out of the $1\sigma$ range of $n_s=0.9682\pm 0.0062$. This conclusion is true just from the ($n_s,r$) curve for the axion inflation. It is also interesting to notice that for a particular $n_s$, with decreasing $f$ reheating temperature increases in accord with the decreasing reheating e-folding number $N_{re}$. Within the $1\sigma$ range our numerical computation shows that $f=6 M_p$ is disfavored as it predicts the maximum value of $n_s^{\rm max}\simeq 0.957$ which outside the $1\sigma$ range of $n_s$ from PLANCK. However for $f =(10,50 M_p)$, we found $n^{\rm max}_s \simeq (0.9644, 0.9655)$  at which  $N_{re}=(1.72,1.3)$. For both the cases, the lowest $n_s \simeq 0.962$ corresponds to the minimum reheating temperature $T_{re}^{\rm min} \simeq (4.9 \times 10^9, 7.6 \times 10^{7}) $ in GeV unit. 
\begin{figure}[t!]
	\centering
	\subfigure[]{\includegraphics[scale=0.45]{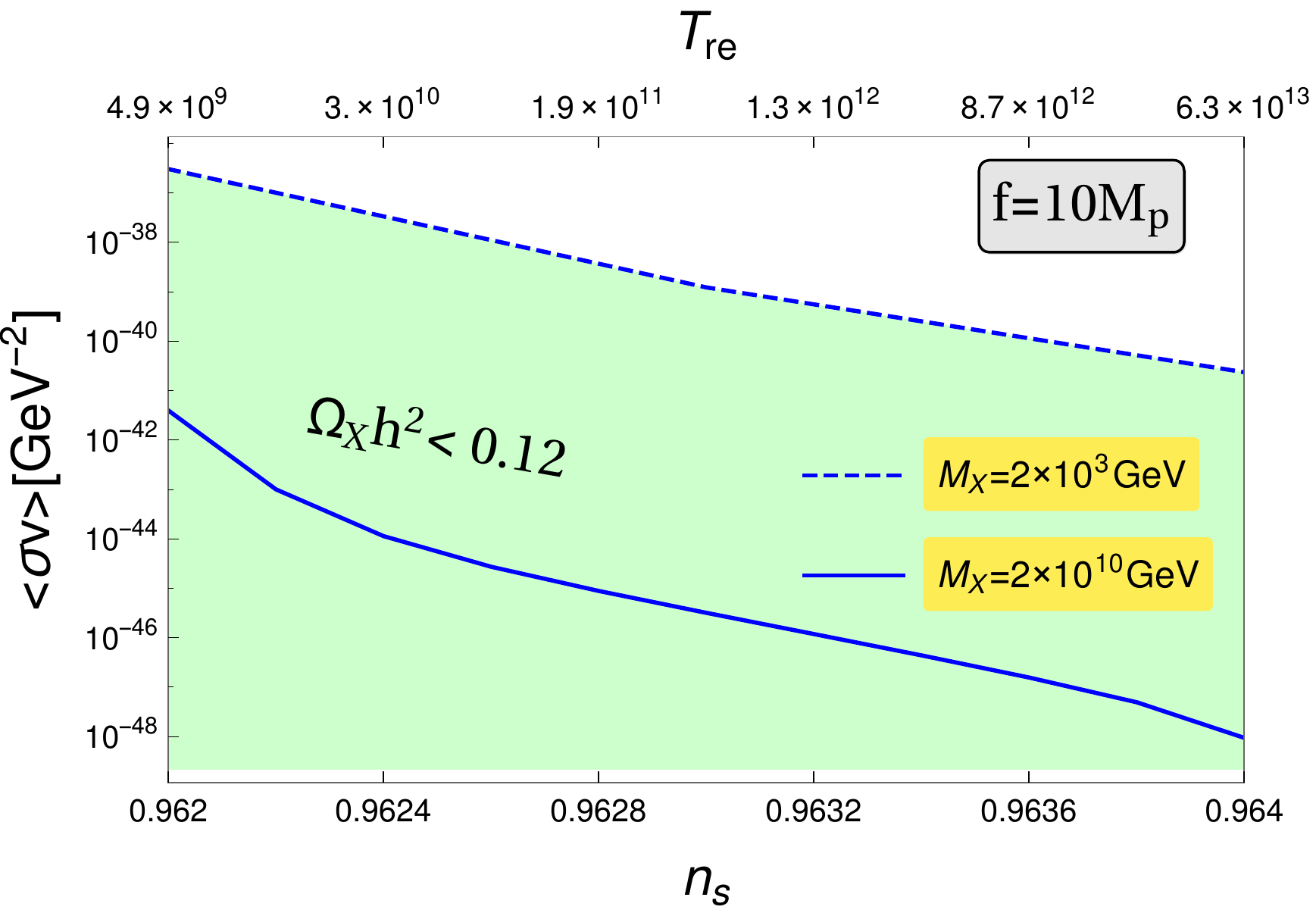}}
	\subfigure[]{\includegraphics[scale=0.45]{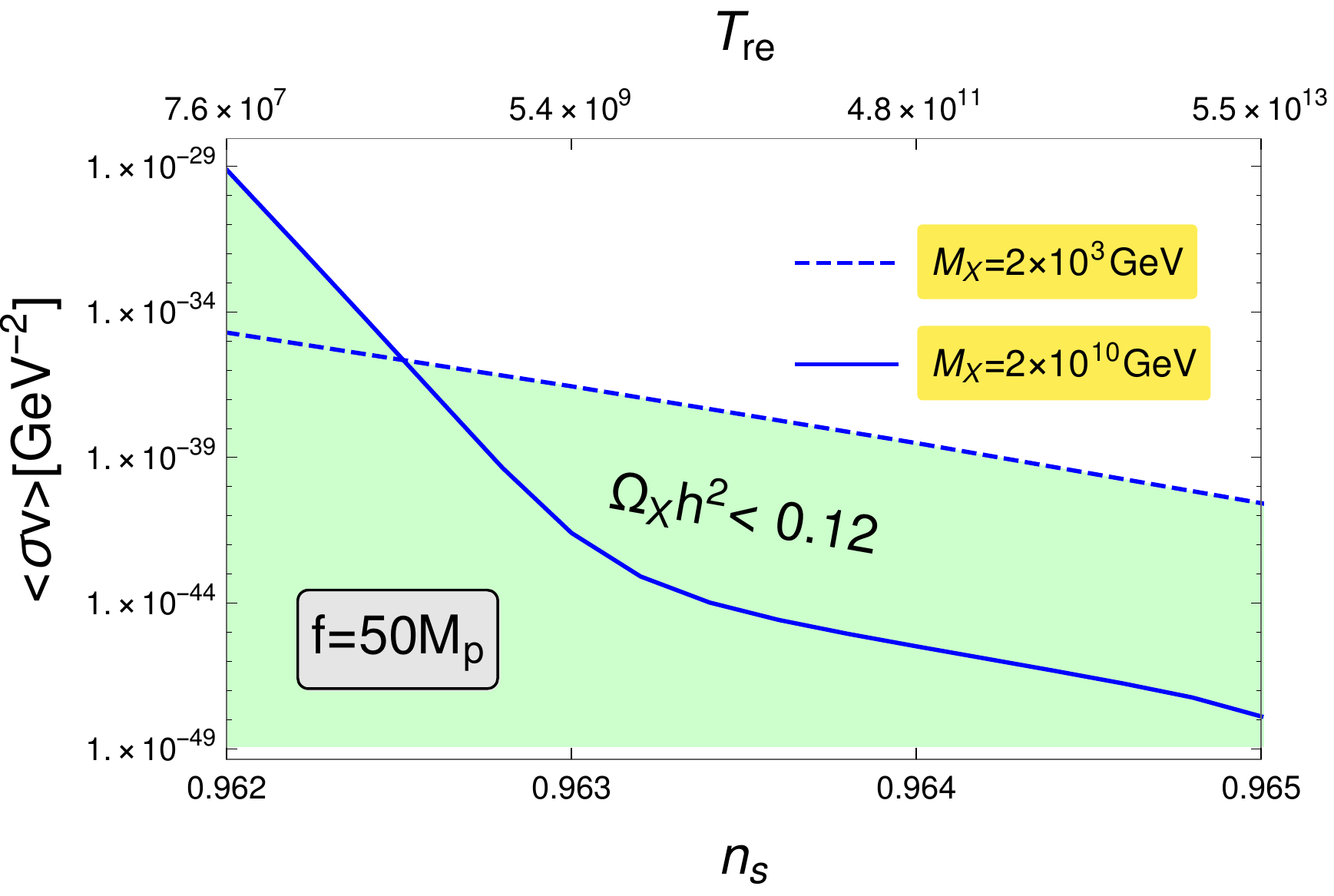}}
	\caption{The same plot as in Fig. \ref{ccs} for the natural inflation model. Axion decay constant for (a) $f=10M_p$ and (b) $f=50M_p$}
	\label{axcs1}
\end{figure}
\begin{figure}[t!]
	\centering
	\subfigure[]{\includegraphics[scale=0.45]{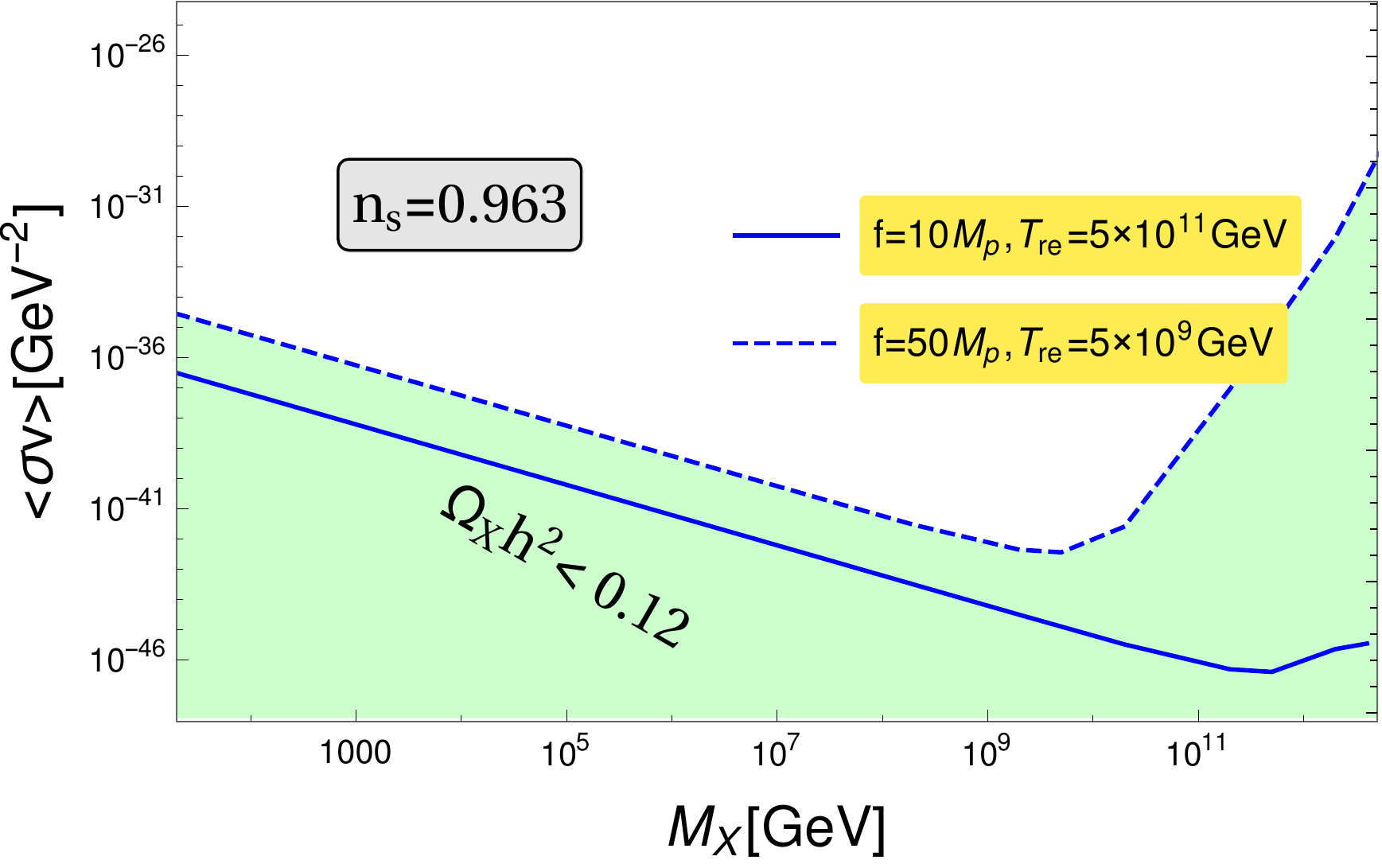}}
	\caption{The same contour plot as in Fig. \ref{ccs} in the $M_X$-$\langle \sigma v\rangle$ space for the natural inflation model. Axion decay constant for (a) $f=10{\rm M_p}$ and (b) $f=50{\rm M_p}$}
	\label{axcs2}
\end{figure}

Now we are in a position to figure out the effect of the axion inflation model in the dark matter phenomenology. In the Figs.\ref{axcs1} and \ref{axcs2}, we have displayed the allowed regions of parameter space for a single component dark matter based on the constraints from CMB observation. The allowed region in $(n_s ~\text{vs}~\left< \sigma v\right>)$ space from the current dark matter relic abundance is shown in fig. \ref{axcs1} for two sample values of dark matter mass $M_X= (2\times 10^3, 2\times 10^9)$ GeV. For $f=10{\rm M_p}$, the reheating temperatures for all the spectral indexes are higher than both the masses and they freeze-in in the radiation dominated era.
However, for $f=50 M_p$ and the dark matter mass $10^9$ GeV, we will have two distinct behaviors given in  Eqs.(\ref{sigmans1}) and (\ref{sigmans2}), which are also reflected in the change of slopes of the contour plots Fig.\ref{axcs1} for $\Omega_Xh^2 =0.12$. The analytic expression for the relic abundance in the different regions can be found in \cite{Giudice:2000ex,Allahverdi:2002pu}. In Fig.(\ref{axcs2}) we present the allowed region in parameter space of $(M_X~ vs~ \left< \sigma v\right>)$ for a fixed value of $n_s$ corresponding to two different reheating temperatures $T_{re} \simeq (5\times 10^{11}, 5\times 10^{9} )$ GeV for two different vales $f=(10,50){\rm M_p}$.
From the physical point of view  as expected for a particular value of $n_s=0.963$, there exists a minimum value of the annihilation cross section $\langle \sigma v\rangle \simeq (5 \times 10^{-47}, 5 \times 10^{-43} ) $ for $f = (10, 50)~M_p$ and $M_X=(5\times 10^{11},5\times 10^{9})$ GeV which are of same order as the reheating temperature. From the physical point of view, this fact can be understood as follows: for dark matter mass $M_X > T_{re}$, the freeze-in temperature $T_{\rm freeze} > T_{re}$, during which the radiation density is very small as most of the energy is in the form of oscillating inflaton field. Therefore, in order to achieve the current dark matter abundance $\Omega_X h^2 \simeq 0.12$ one needs to increase the annihilation cross section as we increase the value of dark matter mass. However, for $M_X < T_{re}$, the freeze-in temperature is obviously $T_{\rm freeze} < T_{re}$,  which is in the radiation dominated phase, and most importantly the radiation temperature $T_{\rm rad}$ becomes inversely proportional to the cosmological scale factor. Therefore, dark matter abundance crucially depends upon the freeze-in time or freeze-in temperature. With the decreasing $M_X$ the freeze-in happens at a late time or, in other words, at a lower value of the freeze-in temperature. This late time freeze-in will  naturally reduce the dark matter abundance. Hence below the reheating temperature, with decreasing $M_X$, one needs to increase cross section $\langle \sigma v\rangle$ in order to produce correct dark matter abundance.

\begin{figure}[t!]
	\centering	
	\subfigure[]{\includegraphics[scale=0.45]{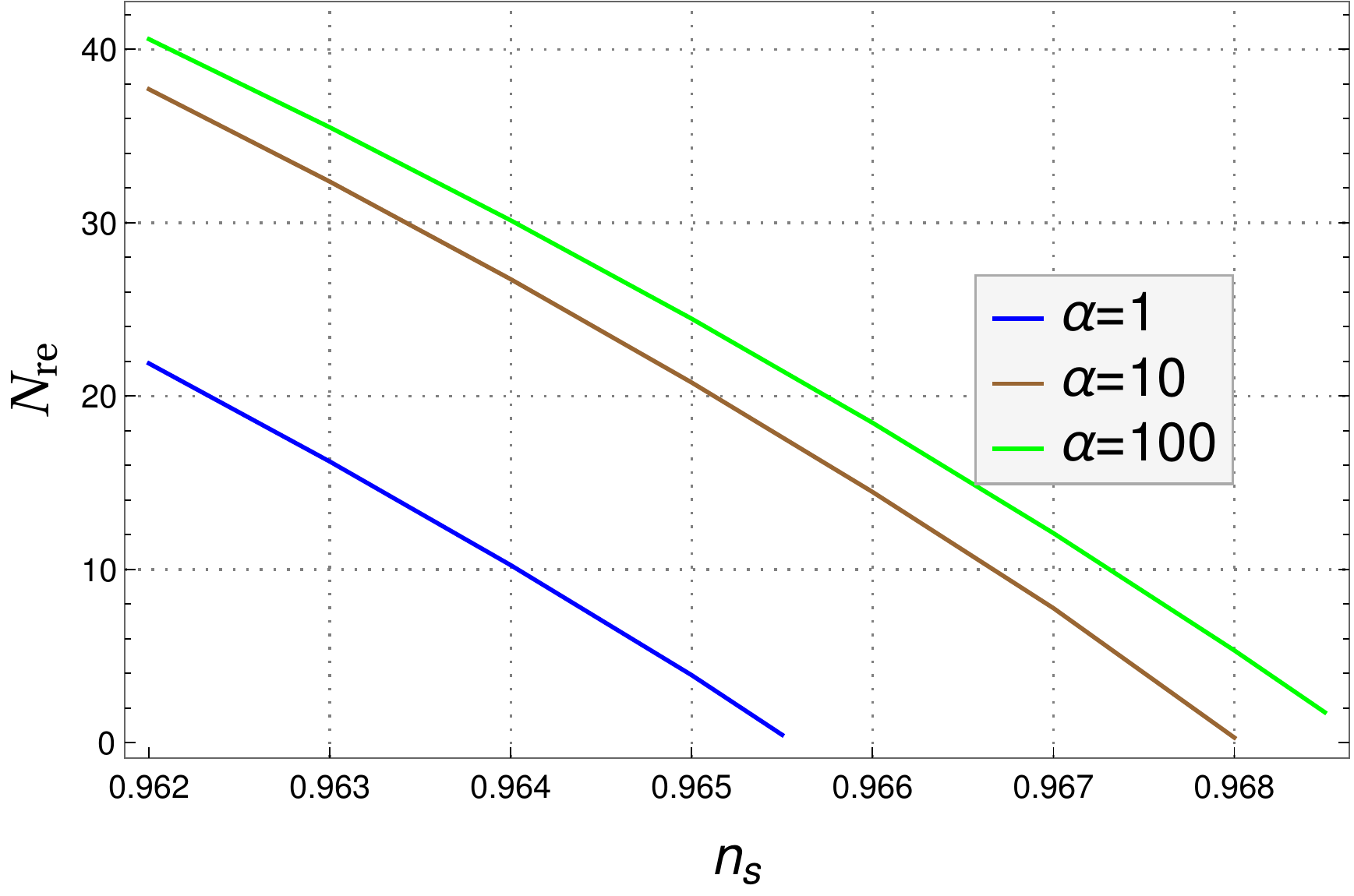}}
	\subfigure[]{\includegraphics[scale=0.451]{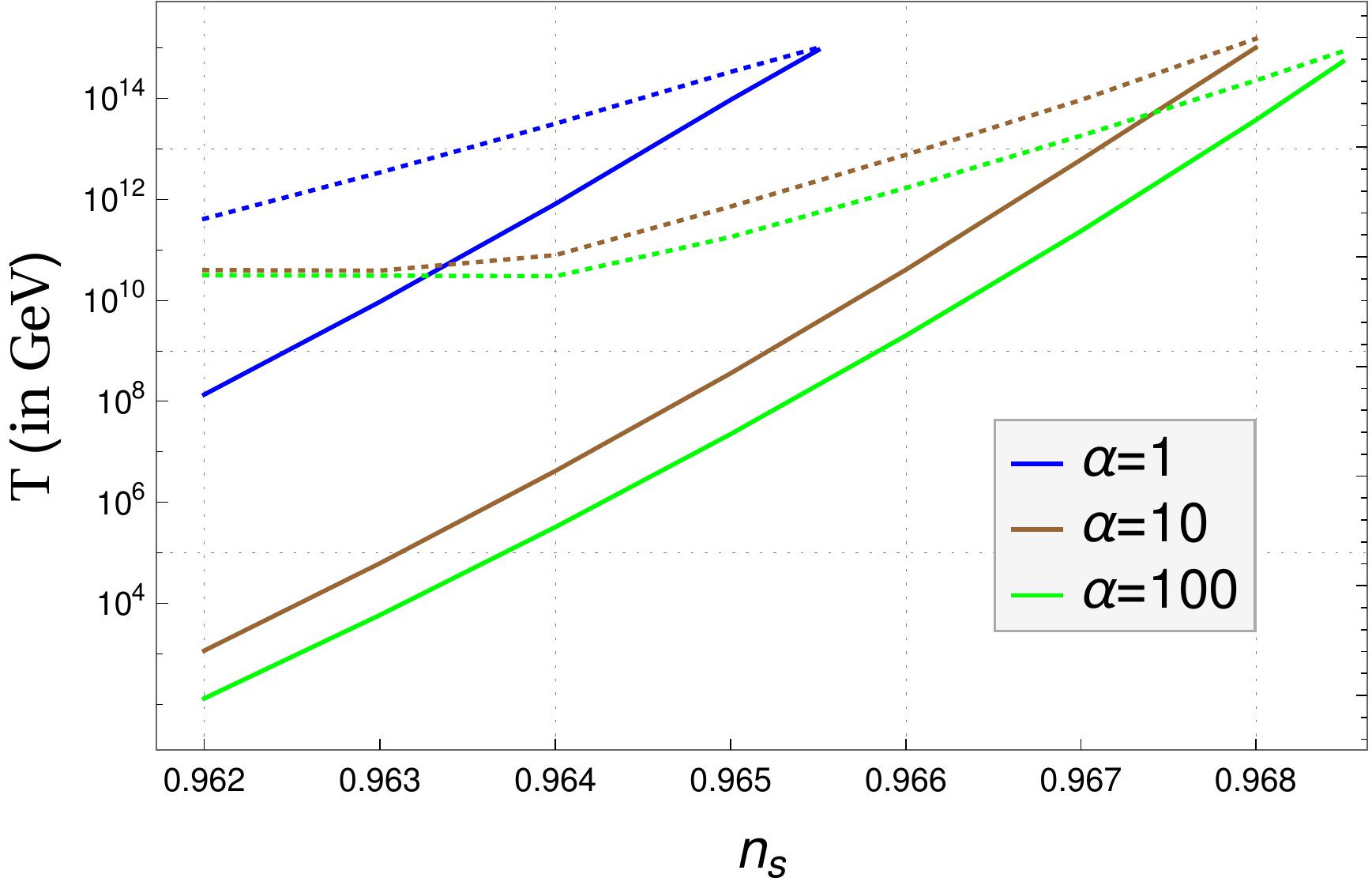}}
	\caption{Variations of (a) $N_{re}$  and (b) $(T_{re}, T_{\max max})$ with respect to $n_s$ have been plotted for $\alpha$-attractor model. We have considered three sample values of $\alpha = (1,10,100)$.}
	\label{alp_nsnret}
\end{figure}
\subsection{Alpha attractor}

In this section will consider a class of models called  $\hyphenB{\alpha-attractor}$ model\cite{Kallosh:2013hoa,Kallosh:2013tua, Kallosh:2013yoa, Ferrara:2013rsa, Ferrara:2013kca, Kallosh:2015lwa} which has recently been proposed to a unify different inflationary models parametrized by a parameter $\alpha$. The uniqueness of this class of models is its conformal property which leads to a universal prediction for the inflationary observables $(n_s,r)$ in favor of Planck observation \cite{PLANCK}. After the conformal transformation of a large class of originally noncanonical inflaton field Lagrangian, one generically gets canonically normalized inflaton field with an exponential potential of the following form

\begin{figure}[t!]
	\centering
	\subfigure[]{\includegraphics[scale=0.45]{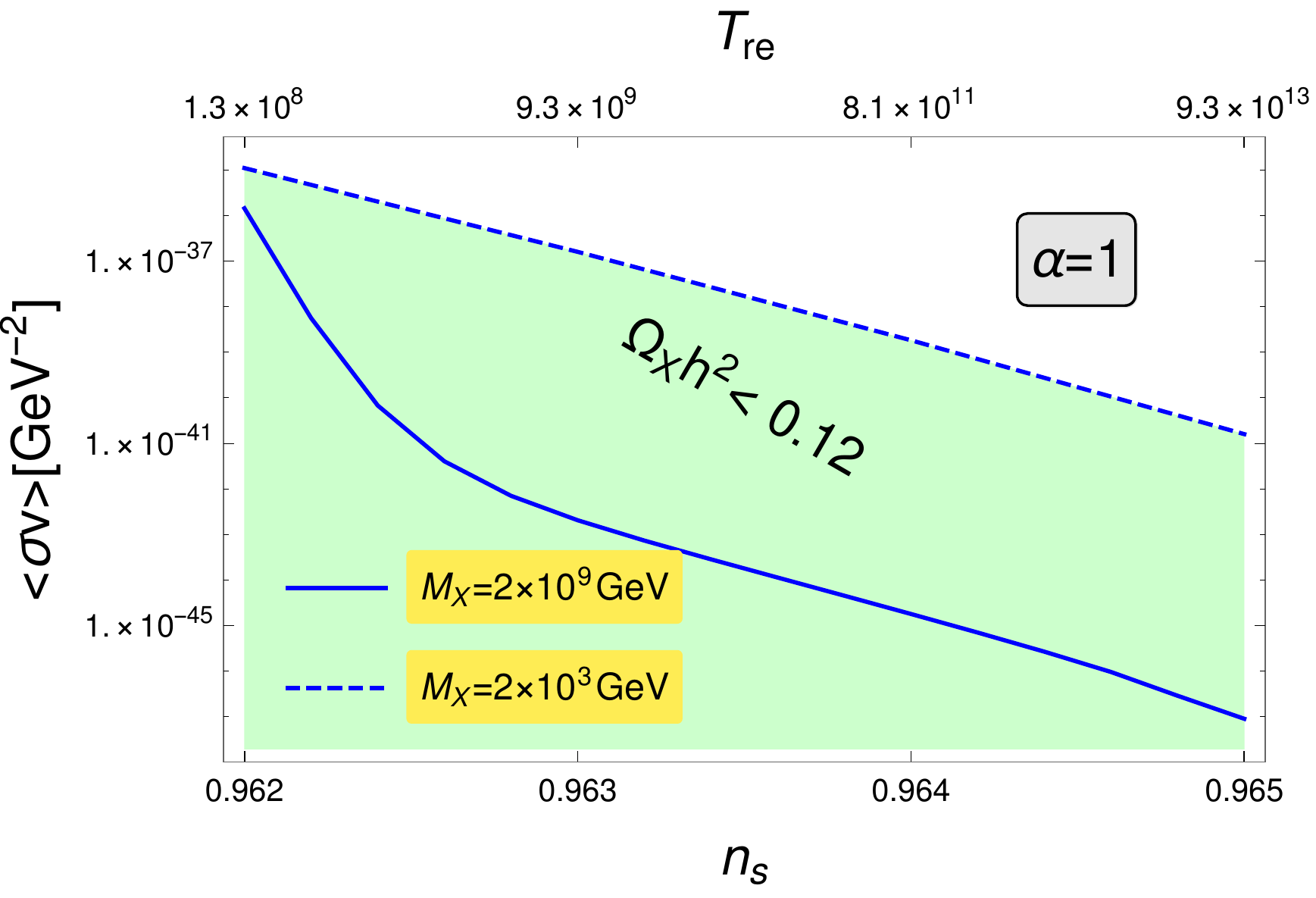}}
	\subfigure[]{\includegraphics[scale=0.454]{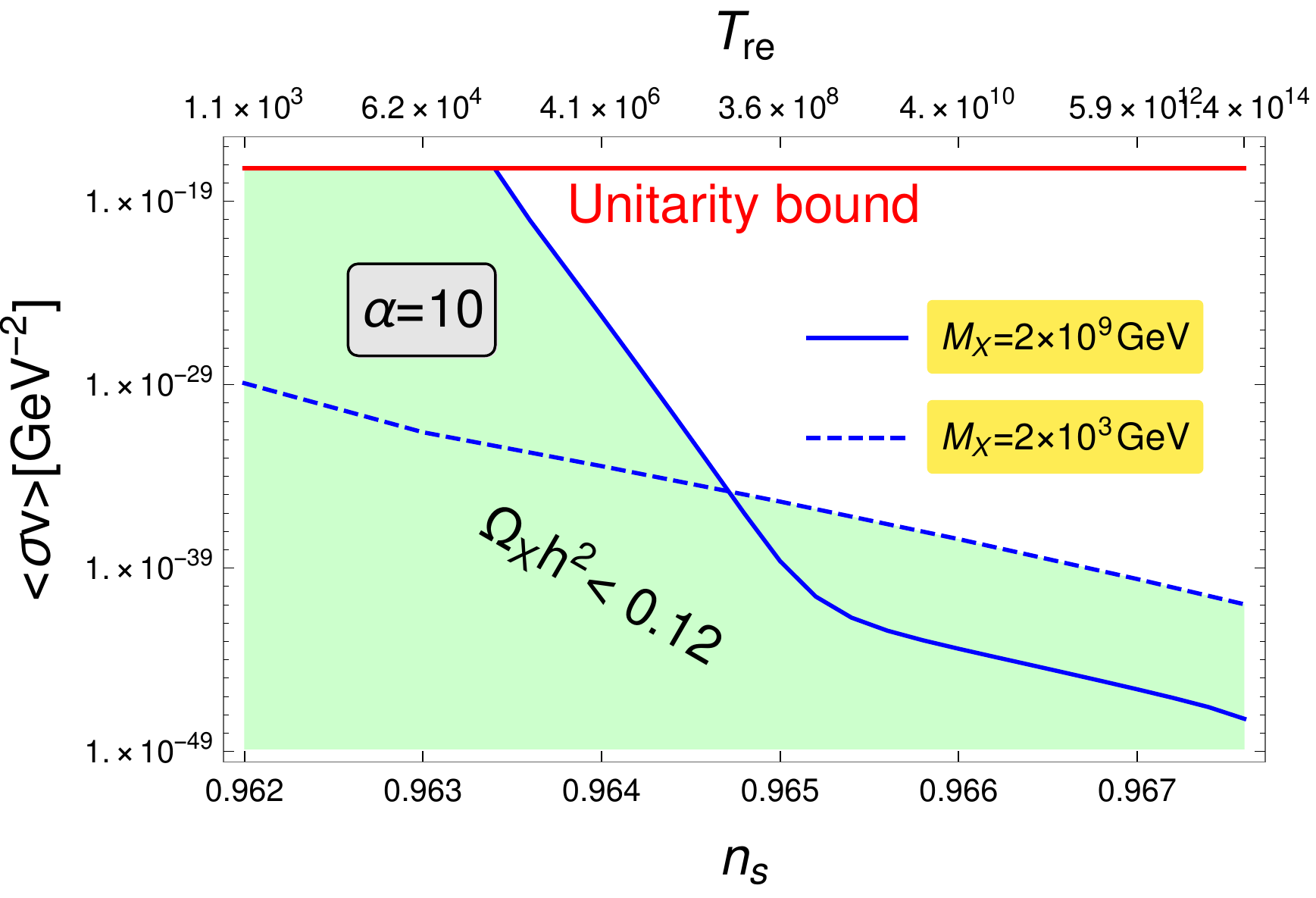}}
	\caption{The shaded region shows the region in the parameter space allowed by current dark matter abundance for two dark matter masses in the $\alpha$-attractor $E$ model. (a) Corresponds to  $\alpha=1$, while Fig.(b) is for $\alpha=10$. .}
	\label{alpcs1}
\end{figure}
\begin{figure}[t!]
	\centering
	\subfigure[]{\includegraphics[scale=0.45]{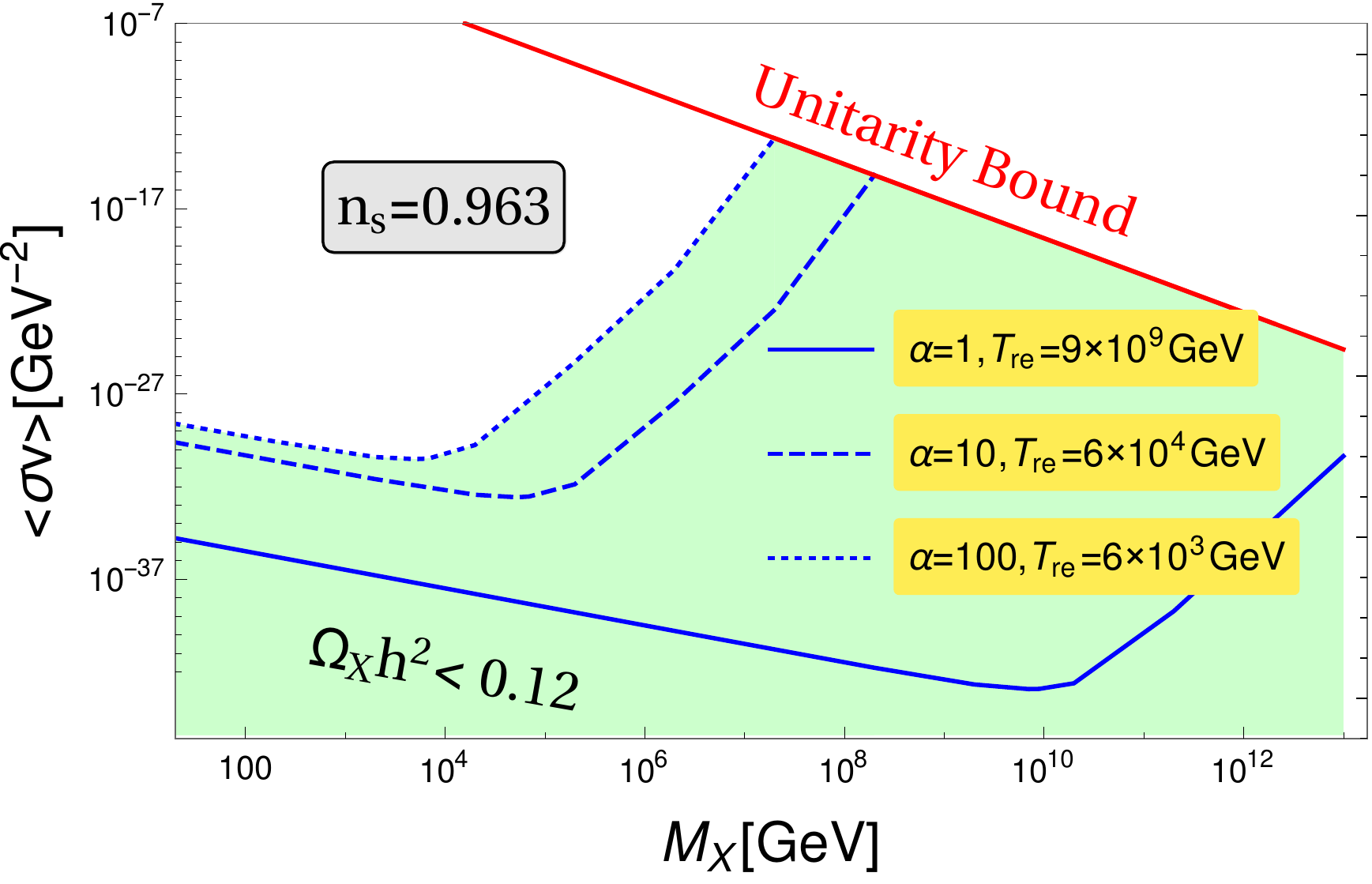}}
	\caption{Considering a sample value of $n_s$ or equivalently $T_{re}$ as given, we plotted $(\langle \sigma v\rangle ~vs~M_X)$ for $\alpha$-attractor $E$ model, for $ \alpha=(1,10,100)$. The solid red line corresponds the unitarity limit $\langle \sigma v\rangle \propto 1/M_X^2$. }
	\label{alpcs2}
\end{figure}

\begin{equation}
V(\phi) = \Lambda^4 \left[  1 - e^{ -     \sqrt{\frac{2}{3\alpha} }    \frac{\phi}{M_p}     } \right]^{2n}
\end{equation}
In the literature, this model is known as the $E$ model. The quantities that we will need for solving the Boltzmann equation is the inflaton equation of state parameter and the inflaton mass, which we will get by expanding the potential around the minimum when $\sqrt{\frac{2}{3\alpha}\phi_{\rm end}}<M_p$ which is equivalent to choosing $\alpha > 0.5n^2$.
\begin{equation}
 V(\phi) \simeq \Lambda^4 \left(\frac{2}{3\alpha}\right)^n \left(\frac{\phi}{M_p}\right)^n
\end{equation}
Now, it is easy to identify the inflaton mass with the tree-level expression as
\begin{equation}
 m_{\phi} = \frac{2\Lambda^2}{\sqrt{3\alpha}M_p}
\end{equation}
and the equation of state parameter, as noted before, is given by $w_{\phi}=0$.

As has been discussed for natural inflation, in this case also we found $\Lambda \simeq 10^{16}$ GeV. The new parameter $\alpha$ determines the shape of the canonically normalized inflaton potential near the minimum. The qualitative behavior of all the plots will be the same as for the other models we have discussed so far. However, the reheating temperature in this class of models can be very small depending on the value of the $\alpha$ parameter. For the purpose of our current study, we have taken $n=1$ and $\alpha = (1,10,100)$ for illustration. It is important to note that $\alpha=1$ encodes two important well studied  inflationary models, namely, Starobinsky\cite{Starobinsky:1980te} and Higgs\cite{Bezrukov:2007ep} inflation. Nonetheless, some important facts can be observed from the Fig.\ref{alp_nsnret} as follows: we clearly see that as one increases the value of $\alpha$, the reheating temperature decreases for a fixed value of $n_s$. For example at $n_s=0.962$ which is the lowest of $1\sigma$ range from PLANCK, we found $T_{re}^{\rm min}\simeq(10^8,10^3,10^2)$ GeV for $\alpha=(1,10,100)$ respectively. The qualitative behavior on the constraints on the dark matter parameter space appeared to be the same as that of the chaotic and natural inflation cases discussed in the previous sections. Specifically, let us emphasize again one of the important results of our analysis shown in Eqs.(\ref{sigmans1}) and (\ref{sigmans2}), which will be satisfied for the $\alpha$-attractor model as well. However, from Figs.(\ref{alpcs1}) and (\ref{alpcs2}), we point out that with increasing $\alpha$, the annihilation cross section increases for a fixed value of the dark matter mass. This fact could be an interesting point to further understand from the theoretical point of view. From our naive numerical solution of Boltzmann equations one finds that for higher value of $\alpha $, the  annihilation cross section could be arbitrarily large depending upon the value of $n_s$ or equivalently the reheating temperature $T_{re}$. However, this should not hold true as the unitarity limit  on $\left<\sigma v\right>_{\rm MAX} = 8\pi/M_X^2$ restricts the allowed region of $n_s$. Therefore, one gets a lower limit on the value of $n_s$ which is coming from the dark matter sector. For example, from Fig.\ref{alpcs1} if one considers $\alpha =10, M_X=2\times 10^9$ GeV, the lowest possible value is $n_s = 0.9634$ set by the unitarity limit(red line). On the other hand, the highest value of the $n_s^{\rm max}\simeq 0.968$ does not depend upon the dark matter parameters as has already been pointed out. This important constraint on the $n_s$ coming from dark matter sector could be very important to understand and needs further study.

\section{Summary and outlook}  
Through our present work the first and the foremost point we wanted to bring to the reader's notice is that it is an important generalization of the work proposed in \cite{Dai:2014jja} by considering explicit decay of inflaton into radiation and dark matter into the reheating constraint analysis.
At this point let us also remind the reader that in all the PLANCK analysis \cite{PLANCK} on constraining the inflationary models, an effective time independent equation of state $w_{\rm eff}$ during reheating is assumed. 
One of the important messages we try convey through the present analysis is that those assumptions have limited applicability. After the inflation, every inflationary model has its own
\begin{figure}
\begin{center}
	\begin{longtable}{>{\fontsize{8pt}{8pt}\selectfont}p{3cm}|>{\fontsize{8pt}{8pt}\selectfont}P{6cm}|>{\fontsize{8pt}{8pt}\selectfont}P{6cm}}
		\caption{Summary of two methods for reheating Constraints}\label{tab:summary} \\
		\hline
		& \textbf{Standard approach\cite{Dai:2014jja,Martin:2014nya}} & \textbf{Our approach} \\
		\hline\hline	
		Assumptions~ &
		\begin{itemize}[label={-},noitemsep,leftmargin=*,topsep=0pt,partopsep=0pt]
			\item During the reheating period time-independent effective equation of state $w_{re}$ is assumed to be a free parameter that parametrizes the expansion of the universe. No microphysics of inflaton decay is considered.
			\item Instantaneous conversion of inflaton energy into radiation.
		\end{itemize} & 
		\begin{itemize}[label={-},noitemsep,leftmargin=*,topsep=0pt,partopsep=0pt]
			\item Reheating phase is described by perturbative inflaton decay into various other fields. Hence $\Gamma_{\phi}$ is a free parameter.
			\item The inflaton equation state is that of the homogeneous inflaton condensate. Hence, total effective equation of state $w_{\rm eff}$ is time-dependent Fig.\ref{weff}. 
		\end{itemize}\\
		Components of the universe &
		\begin{itemize}[label={-},noitemsep,leftmargin=*,topsep=0pt,partopsep=0pt]
			\item Assumes two component universe comprising inflaton and radiation
		\end{itemize} & 
		\begin{itemize}[label={-},noitemsep,leftmargin=*,topsep=0pt,partopsep=0pt]
			\item In principle we can accommodate any number of energy components, such as dark matter and dark radiation and do the analysis.  
		\end{itemize}\\
		Methodology &
		\begin{itemize}[label={-},noitemsep,leftmargin=*,topsep=0pt,partopsep=0pt]
			\item Find out the inflationary quantities $N_{\rm k},r,V_{\rm end},$ etc. in terms of $n_s, A_s$ for a specific inflation model.
			\item Calculate $N_{re}$ in terms of $w_{re}$ using Eq.\ref{nre_dkw}.
			\item Finally one obtains the relation among $T_{re}, n_s$ and $w_{re}$ using Eq.\ref{TreEq}
			\item The inflaton decay constant is indirectly defined through the reheating temperature.
		\end{itemize} & 
		\begin{itemize}[label={-},noitemsep,leftmargin=*,topsep=0pt,partopsep=0pt]
			\item Find out the inflationary quantities $N_{\rm k},r,V_{\rm end},$ etc. in terms of $n_s, A_s$ for a specific inflation model.
			\item Solve the Boltzmann equation considering ($\Gamma_{\phi}, \langle\sigma v\rangle, M_X)$ as a free parameters.
			\item The ``right'' ($\Gamma_{\phi}, \langle\sigma v\rangle$) are uniquely determined by the condition Eq.\ref{bound} for given $(M_X, n_s)$, which are a combination of entropy conservation and background evolution, and dark matter abundance. Inflation fixes the value of $n_s$. Therefore, we only have dark matter mass as a free parameter $M_X$.
		\end{itemize}\\
		\noindent Relations with CMB and primordial density fluctuation&
		\begin{itemize}[label={-},noitemsep,leftmargin=*,topsep=0pt,partopsep=0pt]
			\item With the conventional transfer function connect the primordial spectral tilt with the CMB anisotropy. 
		\end{itemize} & 
		\begin{itemize}[label={-},noitemsep,leftmargin=*,topsep=0pt,partopsep=0pt]
			\item In our present analysis we assumed the conventional relation.
			\item  However our analysis connects dark matter phenomenology with the inflationary observables through reheating. Hence dark matter observation can constrain the inflationary dynamics. 
			\item Therefore, to connect the primordial spectral tilt with the CMB anisotropy appropriate transfer function needs to be derived which explicitly includes the dynamics of reheating.   
		
		\end{itemize}\\\hline
	\end{longtable}
\end{center}
\end{figure}

\begin{figure}[t!]
	\centering
	\subfigure[]{\includegraphics[scale=0.8]{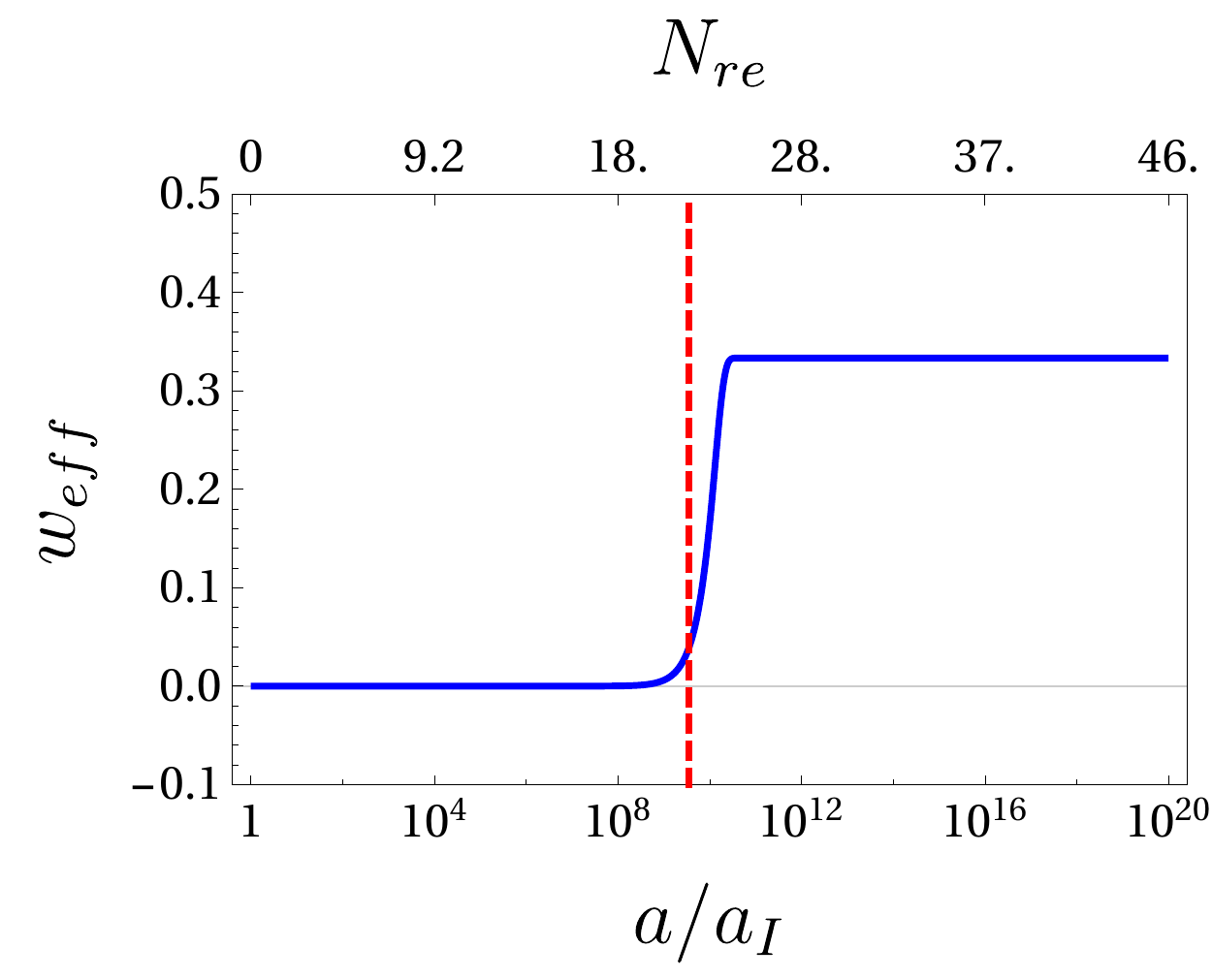}}
	\caption{Variation of effective equation of state $w_{\rm eff} = \left\langle {(3p_{\phi}+\rho_R)}/{3(\rho_{\phi}+\rho_R+\rho_X)} \right\rangle$ during the reheating phase. The vertical red dotted line corresponds to the end of reheating.}
	\label{weff}
\end{figure}

characteristic oscillatory period that contributes to the equation of state during reheating.  
Therefore, considering $w_{\rm eff}$ as a free parameter loses some of the fundamental characteristic properties of the inflaton potential itself. 
Furthermore if reheating occurs for a longer period of time, the time dependent $w_{\rm eff}$ should also be very important to get a precise constraint on any inflationary model. This is where our analysis not only can play an important role in better understanding the inflationary models but also opens up the possibility of understanding the microphysics of the reheating process through CMB physics. To further clarify in the Table-\ref{tab:summary} we summarize and compare our analysis with that of the existing analysis. As we can clearly see, the CMB power spectrum constrains the value of inflation-radiation coupling parametrized by $\Gamma_{\phi}$ through reheating temperature $T_{re}$, which we found to be expressed in terms of spectral index $n_s$ as, 
	\begin{equation}
	{\rm log}\left( T_{re}\right) \propto \left[A + B(n_s - 0.962) + C(n_s - 0.962)^2 \right]. 
	\label{nsTre}
	\end{equation}
	 The usual relation $T_{re} \propto \sqrt{\Gamma_{\phi}} $ will not be exactly correct any more once we consider inflaton decaying into various matter fields. Further in all the previous theoretical as well as PLANCK analysis, complete decay of inflaton is assumed at the beginning of the radiation era. This also cannot be true because of the perturbative part of the reheating. Another interesting point we point out is that irrespective of the model under consideration, our analysis indicates the existence of a universal value of the maximum reheating temperature $T_{re}^{\rm max}\simeq 10^{15}$ GeV and the maximum of value of inflationary e-folding number $N_{\rm max} \simeq 56$ \cite{Maity:2017thw}.

In this work, our main goal was to understand the connection between the CMB anisotropy and the properties of dark matter. Till now the only known quantity related to the dark matter is the dark matter density parametrized by the density parameter $\Omega_X h^2 \simeq 0.12$, which can be extracted from the CMB power spectrum. However in this paper for the first time to our knowledge, we managed to establish the fact that CMB anisotropy not only provides the background value but also can shed light on the microscopic detail of dark matter. In this regard thanks to the Ref.\cite{Dai:2014jja}, a beautiful connection between the CMB power spectrum and the reheating temperature via the inflation has been established. Here we have extended their formalism by including the effect of explicit decay of inflaton into the reheating study \cite{Maity:2017thw}. The main assumption of our analysis was the perturbative decay of inflation. In any inflationary model, the inflaton energy is supposed to be the only source of energy of the current universe. Therefore, in addition to the standard radiation field, we have included the production of a stable dark matter particle species during the reheating period. As has been mentioned in the Introduction, detailed analysis on this has been done in the literature \cite{Chung:1998rq,Allahverdi:2002nb,Allahverdi:2002pu,Pallis:2004yy,Tenkanen:2016twd, Nurmi:2015ema,Bastero-Gil:2015lga, Heikinheimo:2016yds,Kainulainen:2016vzv,Visinelli:2017qga,  Chen:2017kvz, Enqvist:2017kzh, DEramo:2017ecx} without any constraint from the CMB. However, let us emphasize again that we reanalyzed the dark matter production considering the important constraints coming from observed CMB anisotropy.   
 
Other important conclusions of our analysis is that for a particular inflation model, the inflationary scalar spectral index that is directly connected with the CMB power spectrum can uniquely fix the dark matter parameter space $(M_X, \langle \sigma v\rangle)$, through the following important relations for different dark matter mass ranges,  
  \begin{eqnarray} 
 &&\left< \sigma v \right> \Big|_{M_X>T_{re}} \propto 10^{-7A -7 B(n_s - 0.962) - 7 C(n_s - 0.962)^2} .\nno\\
&&  \left< \sigma v \right> \Big|_{M_X<T_{re}} \propto 10^{-A - B(n_s - 0.962) - C(n_s - 0.962)^2} .
  \end{eqnarray} 
As is clear from the above expressions for the dark matter annihilation cross section, which turned out to be very sensitive to the inflationary scalar spectral index because of the power-law form, it is very important to pinpoint the value of $n_s$ in the future CMB experiments. It is clear from the expression that for a given dark matter mass and the inflationary model, the dark matter scattering cross section will be within the bound coming from the $2\sigma$ error bar on   $n_s =0.9670 \pm 0.0074 $ from Planck and BK14 and BAO data. For marginally relevant axion inflation models, we found for axion decay constant $f=10 M_p$ and the dark matter mass, $M_X = 10^3 \mbox{GeV}$, the dark matter cross-section should be with $10^{-39} > \langle \sigma v \rangle > 10^{-41} \mbox{GeV}^{-2}$ which is very narrow within the $2\sigma$ error of $n_s$ mentioned before. If we consider one of the observationally favorable models of $\alpha$ attractor with $\alpha =10$, we get a large range of annihilation cross section $10^{-29}> \langle\sigma v\rangle > 10^{-42}$ $\mbox{GeV}^{-2}$ possible for dark matter mass $M_X=10^3$ GeV. More details of this bound on the allowed ranges of $\langle \sigma \rangle$ for different mass range has already been discussed in the main text considering various models of inflation.

 Explicit model building in the dark matter sector during the reheating period could be an important research direction.   
In addition to the connection we have been discussing, we also found that to satisfy the bound on the current dark matter abundance, freeze-in is the only mechanism through which dark matter with $M_X \gg T_{re}$ can be produced. Our numerical analysis also showed that dark matter production during reheating does not significantly affect the determination of reheating temperature. 

In the present analysis, we have only considered the homogeneous evolution. It would be of utmost importance to analyze the evolution of perturbations of radiation and dark matter components and study their spectral properties which can give further constraints on our parameters. Most importantly in our analysis inflation and the subsequent reheating control the dynamics of all the energy components such as radiation and dark matter of our universe. Reheating is effective in the subhorizon scale. Therefore any small-scale observables related to CMB and matter distribution could play important role in constraining inflationary models though our analysis. One of the important such set of observables could be the well-known small-scale $\mu$-type and $y$-type spectral distortions of CMB. The standard $\Lambda$CDM cosmology already predicts those spectral distortions through standard photon-charge particle interaction \cite{spectral1,spectral2} at different redshift values. However, at present those distortion parameters are tightly constrained by COBE and FIRAS experiments, $|\mu| < 9 \times 10^{−5}$ and $ y < 1.5 \times 10^{−5}$ \cite{cobe}. However, future projected sensitivity of those quantities in new experiments like PIXIE \cite{pixie} and PRISM \cite{prism} are within $ 10^{-8}- 10^{-9}$. Therefore, it would be important to understand various physical processes that can give rise to any deviation from a black body spectrum. In our present analysis, we consider the scenario where the energy is being extracted out of the radiation to dark matter and depending upon the dark matter mass and the inflationary scalar spectral index, the freezing out of dark matter happens in a large range of cosmological redshift values. Therefore, this energy extraction process can leave its footprint in the CMB spectral distortion parameters \cite{spectral2,spectral3}, which can further constrain the inflationary models. We leave these important topics for our future studies.

An important assumption in our analysis that needs further investigation is the assumption of the perturbative decay of inflaton during reheating. The perturbative decay of inflaton \cite{Dolgov:1982th, Abbott:1982hn} has been parametrized by an effective phenomenological friction term with inflaton  decay constant $\Gamma_{\phi}$. However, from the action principle, this is very difficult to generate. Therefore, as has been mentioned before, one should construct an explicit dark matter model. Most importantly it has long been argued that the nonperturbative decay of inflaton will be very important and efficient at the initial stage of the reheating phase. In the literature this phase is known as preheating \cite{Kofman:1994rk,Shtanov:1994ce,Kofman:1997yn, Bassett:2005xm, Amin:2014eta}. However, once the amplitude of the oscillating inflaton is small after preheating, the perturbative decay will automatically come into play. 
 Hence, it would be more appropriate to understand the nonperturbative dynamics and how it sets the initial conditions for the perturbative reheating where our analysis will be important. This subject is beyond the scope of our present work and will be addressed in a future publication. Nonetheless, as long as the coupling parameters are such that perturbative decay is the only way to reheat the universe, all our conclusions will be qualitatively correct.

To the end let us elaborate one more issue which we have already mentioned in the last point of the Table \ref{tab:summary}. The issue is related to the relation between the CMB anisotropy and primordial anisotropy originated from the quantum fluctuation of the inflaton field. The evolution of the primary power spectrum of the CMB is generally determined through transfer function. This transfer function, which is intimately related to the Sachs-Wolfe effect \cite{sachs-wolfe} entails a simple geometrical scaling relation. Furthermore, there exists an inherent connection between this aforementioned scaling relation and the well-known geometrical parameter degeneracy in determining the CMB spectra [\cite{degeneracy}, {\it and references therein}]. The parameter degeneracy states that the same anisotropy spectrum can be produced even if cosmological constant and spatial curvature is varied keeping the size of the last scattering surface constant. In the usual analysis of this transfer function, the initial condition for the perturbation is set at the BBN. However the initial spectral density distribution at the BBN for various matter components should originate from primordial spectrum through the evolution during the intermediate reheating phase. In our present analysis we solved the homogeneous Boltzmann equations for all the important energy components of our universe starting from the end of inflation, and we tried to understand the constraints on inflation supplemented by not only CMB but also the dark matter abundance. By this we can establish a direct connection among the inflationary dynamics, CMB anisotropy and dark matter phenomenology the via reheating phase. Therefore, to have a complete correspondence between the CMB and the primordial anisotropy, we need to have an additional transfer function that can connect the anisotropy at the end of inflation and the end of reheating.
In order to find out that additional transfer function one needs to solve inhomogeneous Boltzmann equations for various components during the reheating phase. For for those equations to be solved, inflationary dynamics provides us precise initial conditions at the end of inflation. In this additional phase a new parameter degeneracy may appear or if we include the dynamical generation of cosmological constant from the inflaton during reheating, it may lift some amount of  degeneracy in the transfer function. All these important questions we leave for our future studies.

\section{Acknowledgment}

We are very thankful for numerous vibrant discussions with our HEP and GRAVITY group members. We thank Kazunori Kohri for useful comments on the draft. We gratefully acknowledge very constructive suggestions from the anonymous referee, which immensely helped our understanding and improved our presentation.

\end{document}